\documentstyle[aps,pra,psfig,epsf,epsfig,twocolumn]{revtex} 

\newcommand{\ints}{\int_0^{\hspace{.5ex}\infty}
{^{\hspace{-2.8ex}\textstyle '}}\hspace{2ex}}
\newcommand{\intss}{\int_0^{\hspace{1ex}\infty}
{^{\hspace{-3.5ex}\textstyle ''}}\hspace{2ex}}
\newcommand{\Ints}{\int_0^{\hspace{.5ex}\infty}
{^{^{^{\hspace{-2.8ex}\textstyle '}}\hspace{1.5ex}}}}
\newcommand{\Intss}{\int_0^{\hspace{1ex}\infty}
{^{^{^{\hspace{-3.5ex}\textstyle ''}}\hspace{1.5ex}}}}
\newcommand{\sumprime}{\hspace{-.5ex}{^{\textstyle'}}}

\begin{document}
\draft

\title{Resonant dipole-dipole interaction in the presence of
dispersing and absorbing surroundings}

\author{Ho Trung Dung\cite{byline},
Ludwig Kn\"{o}ll, and Dirk-Gunnar Welsch}
\address{
Theoretisch-Physikalisches Institut,
Friedrich-Schiller-Universit\"{a}t Jena,
Max-Wien-Platz 1, 07743 Jena, Germany}

\date{May 10, 2001}
\maketitle

\begin{abstract}
Within the framework of quantization of the macroscopic electromagnetic
field, equations of motion and an effective Hamiltonian for treating
both the resonant dipole-dipole interaction between two-level
atoms and the resonant atom-field interaction are derived, which can
suitably be used for studying the influence of arbitrary dispersing
and absorbing material surroundings on these interactions.
The theory is applied to the study of the
transient behavior of two atoms that initially
share a single excitation, with special emphasis on
the role of the two competing processes of virtual and real photon
exchange in the energy transfer between the atoms.
In particular, it is shown that for weak atom-field interaction
there is a time window, where the energy transfer follows a
rate regime of the type obtained by ordinary second-order
perturbation theory. Finally, the resonant dipole-dipole interaction
is shown to give rise to a doublet spectrum of the emitted light for
weak atom-field interaction and a triplet spectrum for
strong atom-field interaction.
\end{abstract}


\narrowtext


\section{Introduction}
\label{Intro}

Recently, several implementations of quantum logic gates relying
on the resonant dipole-dipole interaction between two atomic qubits
have been proposed. The atomic qubits could be impurity atoms in the
condensed phase \cite{Lukin00} or quantum dots embedded in a
semiconductor \cite{Barenco95}. Here and for many other
applications in practice, the question of the influence of real
material surroundings on the mutual interaction of the dipoles
arises. Moreover, tailor-made material surroundings offer the
possibility of controlling the mutual interaction of the dipoles.
In particular, non-absorbing photonic crystals \cite{John91} and
non-absorbing Fabry-P\'erot cavities
\cite{Kobayashi95,Takada99}
have been studied within the framework of mode expansion.
Effects of material dispersion and absorption have been
taken into account for bulk material \cite{Juzeliunas94},
Fabry-P\'erot cavities \cite{Agarwal98}, and
microspheres \cite{Ho01a}.

In the regime of weak atom-field coupling,
the mutual resonant interaction of atoms
has typically been characterized by an effective two-body
potential involving atomic variables only. By analyzing a
one-dimensional
cavity model, it has been shown that this concept may fail
to give a correct description of the interaction at least
in the strong-coupling regime, where the electromagnetic
field degrees of freedom can no longer be eliminated \cite{Goldstein97}.
In that case, it may be instructive to treat the interaction of the
atoms with the on-resonant part of the electromagnetic field
exactly and the interaction with the off-resonant part in a
perturbative manner. In particular, for two two-level atoms that
resonantly interact with cavity-type modes, an effective
Hamiltonian of the form
\begin{eqnarray}
\label{e1}
\lefteqn{
       \hat{H}_{\rm eff} =
      \hbar\sum_\lambda \omega_\lambda
      \hat{a}^\dagger_\lambda\hat{a}_\lambda
      + \hbar\left(\omega_{u_A} \hat{\sigma}^\dagger_A\hat{\sigma}_A
      + \omega_{u_B} \hat{\sigma}^\dagger_B\hat{\sigma}_B\right)
}
\nonumber\\&&\hspace{7ex}
     +\,\hbar M\left( \hat{\sigma}^\dagger_A\hat{\sigma}_B
       +  \hat{\sigma}_A\hat{\sigma}^\dagger_B\right)
\nonumber\\[1ex]&&\hspace{7ex}
     +\,\hbar\sum_\lambda\left(
      \kappa_{A\lambda}\hat{a}_\lambda \hat{\sigma}^\dagger_A
      + \kappa_{B\lambda}\hat{a}_\lambda \hat{\sigma}^\dagger_B
      +{\rm H.c.}\right)
\end{eqnarray}
has been proposed \cite{Kurizki96}. Here, the first term is the
energy of the cavity modes [of frequencies $\omega_\lambda$
and photon destruction (creation) operators $\hat{a}_\lambda$
$(\hat{a}_\lambda^\dagger)$], the second term is the energy
of the two atoms, where $\hat{\sigma}_{A(B)}$
and $\hat{\sigma}_{A(B)}^\dagger$ are the Pauli operators of the
two-level atom $A(B)$, and $\hbar\omega_{u_{A(B)}}$ is the
upper-state energy, the third term is the energy of the so-called
resonant dipole-dipole interaction of the atoms to each other
($\hbar M$ - coupling
strength), and the fourth term is the familiar interaction energy
between the atoms and the cavity modes in the rotating-wave
approximation. Apart from the fact that the coupling parameters in
the Hamiltonian (\ref{e1}) are not specified and their relation
to each other thus remains unclear (also see
\cite{Kurizki88,Zheng96}), material absorption cannot be
taken into account, because of the underlying concept of mode
decomposition.

In this paper we give a rigorous derivation of an effective
Hamiltonian, which -- although at first glance looks like that
in Eq.~(\ref{e1}) -- applies to atoms in arbitrary dispersing
and absorbing material surroundings, with the coupling parameters
being well defined. For this purpose, we start from the
multipolar Hamiltonian governing the motion of the coupled system of
two-level atoms and the medium-assisted electromagnetic field,
with the medium being described in terms of a spatially varying permittivity
that is a complex function of frequency (for a review,
see \cite{Knoll01}). The quantity that essentially governs the
strength of the atom-field interaction is the (classical) Green
tensor of the inhomogeneous Helmholtz equation with the
space- and frequency-dependent complex permittivity of
the material surroundings.
Treating the off-resonant part of the
atom-field interaction in a (coarse-grained) Markov approximation,
whereas the on-resonant part of the atom-field interaction is left
in the original form, we derive the density-matrix equations of
motion for the coupled atom-field system, where the off-resonant
atom-field interaction is eliminated in favor of the resonant
dipole-dipole interaction. We then show that under
certain conditions an effective Hamiltonian that governs the
motion can be constructed. The results show that the
resonant dipole-dipole interaction (via the off-resonant
atom-field interaction) and the on-resonant atom-field
interaction, respectively, are essentially determined by the
real and the imaginary part of the medium-assisted Green
tensor. Whereas the spontaneous decay is only determined
by the imaginary part, the mutual interaction of the atoms
is determined by both the real and the imaginary part of the
Green tensor.

To illustrate the theory, we examine, for
both weak and strong atom-field interaction, the temporal evolution
of two two-level atoms that initially share a single excitation,
with special emphasis on the interatomic energy transfer.
In particular for weak atom-field interaction, from
the exact time dependence of the upper-state population of the
acceptor atom we infer an energy transfer rate. We show that
it is essentially the same rate as the commonly used rate, i.e.,
the transition probability per unit time which is
obtained by means of Fermi's golden rule in second-order
perturbation theory with regard to the original Hamiltonian.
Finally we address the problem of the influence of the resonant
dipole-dipole interaction on the spectrum of the light
emitted by the atoms in the limits of weak and strong
atom-field interaction.

The paper is organized as follows. In Section \ref{densmatrix}
the density-matrix equations of motion for the system that
consists of the atoms and the resonant part of the electromagnetic
field are derived. The problem of deriving them from an
effective Hamiltonian is studied in Section \ref{effham}.
The temporal evolution
of two atoms that initially share a single excitation is
studied in Section \ref{sysdy}. The problem of determining
an energy transfer rate is considered in Section \ref{rateregime},
and Section \ref{spectrum} is devoted to the spectral properties
of the emitted light. Finally, a summary and some concluding remarks
are given in Section \ref{concl}.


\section{Density-matrix equation}
\label{densmatrix}

Let us consider $N$ two-level atoms [positions ${\bf r}_A$, transition
frequencies $\omega_A$, dipole moments ${\bf d}_A$
($A$ $\!=$ $\!1,2,...,N$)] that resonantly interact with the
electromagnetic field via electric-dipole transitions
in the presence of dispersing and absorbing bodies.
The corresponding multipolar-coupling Hamiltonian reads as
\cite{Knoll01,Ho01b}
\begin{eqnarray}
\label{e2}
\lefteqn{
   \hat{H} = \int \!{\rm d}^3{\bf r}
   \! \int_0^\infty \!\!{\rm d}\omega \,\hbar\omega
   \,\hat{\bf f}^\dagger({\bf r},\omega){}\hat{\bf f}({\bf r},\omega)
    + \sum_{A} {\textstyle{1\over 2}}\hbar\omega_A \hat{\sigma}_{Az}
}
\nonumber\\[1ex]&&\hspace{10ex}
    - \sum_{A} \int_0^\infty {\rm d}\omega \left[
    \hat{\bf d}_A
    \underline{\hat{\bf E}}({\bf r}_A,\omega)
    + {\rm H.c.}\right],
\end{eqnarray}
where
\begin{equation}
\label{e2.1}
        \hat{\bf d}_A = {\bf d}_A\hat{\sigma}_A
        + {\bf d}_A^\ast \hat{\sigma}_A^\dagger
\end{equation}
and
\begin{equation}
\label{e3}
     \underline{\hat{\bf E}}({\bf r},\omega)
      \!=\! i \sqrt{\frac{\hbar}{\pi\varepsilon_0}}
     \frac{\omega^2}{c^2}
   \!\!\int \!\!{\rm d}^3{\bf r}'
   \sqrt{\varepsilon_{\rm I}({\bf r}',\omega)}
   \,\bbox{G}({\bf r},{\bf r}',\omega)
   {}\hat{\bf f}({\bf r}',\omega).
\end{equation}
Here, $\hat{\bf f}({\bf r},\omega)$ and
$\hat{\bf f}^\dagger({\bf r},\omega)$ are bosonic fields which
play the role of the fundamental variables of the electromagnetic
field and the medium, including a reservoir necessarily associated
with the losses in the medium, $\bbox{G}({\bf r},{\bf r}',\omega)$
is the classical Green tensor, and
\mbox{$\varepsilon({\bf r},\omega)$
$\!=$ $\!\varepsilon_{\rm R}({\bf r},\omega)$
$\!+$ $\!i\varepsilon_{\rm I}({\bf r},\omega)$} is
the complex (Kramers-Kronig consistent)
permittivity. It should be pointed out that there are no direct
Coulomb forces between particles in the Hamiltonian (\ref{e2}); all
interactions are mediated by the medium-assisted electromagnetic field.

With regard to the interaction of the atoms with the
electromagnetic field, it is convenient to decompose the latter
into an on-resonant part (denoted by $\ints {\rm d\omega}\ldots$)
and an off-resonant part (denoted by$\intss {\rm d\omega}\ldots$).
Let us now consider the temporal evolution of the system that consists
of the atoms and the on-resonant part of the electromagnetic field.
If $\hat{O}$ is any system operator, we may write its equation of motion in
the Heisenberg picture as, on recalling the
Hamiltonian (\ref{e2}),
\begin{eqnarray}
\label{e3.1b}
\lefteqn{
       \,\dot{\!\hat{O}} = -{i\over \hbar}
       \left[\hat{O},\hat{H}_{\rm S}\right]
}
\nonumber\\[1ex]&&\hspace{5ex}
       +\,{i\over \hbar} \sum_A \Intss {\rm d}\omega\,
       \left\{\left[\hat{O},\hat{\bf d}_A\right]
       \underline{\hat{\bf E}}({\bf  r}_A,\omega)
       \right.
\nonumber\\[1ex]&&\hspace{20ex}
       \left.
       +\,\underline{\hat{\bf E}}^\dagger({\bf  r}_A,\omega)
       \left[\hat{O},\hat{\bf d}_A\right]\right\},
\end{eqnarray}
where
\begin{eqnarray}
\label{e3.1a}
\lefteqn{
   \hat{H}_{\rm S} = \int \!{\rm d}^3{\bf r}
   \! \Ints \!\!{\rm d}\omega \,\hbar\omega
   \,\hat{\bf f}^\dagger({\bf r},\omega){}\hat{\bf f}({\bf r},\omega)
    + \sum_{A} {\textstyle{1\over 2}}\hbar\omega_A \hat{\sigma}_{Az}
}
\nonumber\\[1ex]&&\hspace{15ex}
    - \sum_{A} \Ints {\rm d}\omega \left[
    \hat{\bf d}_A
    \underline{\hat{\bf E}}({\bf r}_A,\omega)
    + {\rm H.c.}\right].
\end{eqnarray}
Note that in Eq.~(\ref{e3.1b}) normal ordering
is adopted such that $\underline{\hat{\bf E}}({\bf  r}_A,\omega)$
is on the right-hand side and
$\underline{\hat{\bf E}}^\dagger({\bf  r}_A,\omega)$
is on the left-hand side in operator products.
According to Eq.~(\ref{e3}), the operators $\underline{\hat{\bf E}}
({\bf r}_A,\omega)$ and $\underline{\hat{\bf E}}^\dagger
({\bf r}_A,\omega)$ in Eq.~(\ref{e3.1b}) are thought to be expressed
in terms of the basic-variable operators $\hat{\bf f}({\bf r},\omega)$ and
$\hat{\bf f}^\dagger({\bf r},\omega)$ respectively.
It is not difficult to see that $\hat{\bf f}({\bf r},\omega)$
obeys the Heisenberg equation of motion
\begin{eqnarray}
\label{e3.1}
\lefteqn{
       \dot{\hat{\bf f}}({\bf r},\omega)
       = -i\omega \hat{\bf f}({\bf r},\omega)
}
\nonumber\\[1ex]&&\hspace{10ex}
       +\, {\omega^2\over c^2}
       \sqrt{\varepsilon_{\rm I}({\bf r},\omega)
       \over  \hbar\pi\varepsilon_0}
       \sum_{A} \hat{\bf d}_{A}
       \bbox{G}^\ast({\bf r}_{A},{\bf r},\omega).
\end{eqnarray}
We now solve Eq.~(\ref{e3.1}) formally, insert the result into
Eq.~(\ref{e3.1b}), apply a (coarse-grained)
Markov approximation to the slowly varying atomic variables in
the time integrals in the off-resonant frequency integrals,
and assume that the off-resonant free-field is initially
\mbox{($t$ $\!=$ $\!0$)}
prepared in the vacuum state. After some algebra,
we arrive at the following equation of motion
for the expectation value of the system operator $\hat{O}$
(Appendix \ref{appA}):
\begin{eqnarray}
\label{e3.1c}
\lefteqn{
       \bigl\langle\, \dot{\!\hat{O}} \bigr\rangle
       = -{i\over \hbar}
       \left\langle \left[\hat{O},\,\hat{\!\tilde{H}}_{\rm S}
       \right] \right\rangle
}
\nonumber\\[1ex]&&\hspace{0ex}
       +\, i\sum_{A,A'}\sumprime
       \left\{
       \delta^-_{A^\ast A'}
       \left\langle
       \left[\hat{O},\hat{\sigma}^\dagger_A\right]\hat{\sigma}_{A'}
       \right\rangle
        +\delta^+_{A A^{'\ast}}
	\left\langle
        \left[\hat{O},\hat{\sigma}_A\right]\hat{\sigma}^\dagger_{A'}
        \right\rangle
        \right.
\nonumber\\[1ex]&&\hspace{2ex}
        \left.
        +\,\delta^-_{A A^{'\ast}}
	\left\langle \hat{\sigma}^\dagger_{A'}\left[\hat{O},
        \hat{\sigma}_A\right]\right\rangle
        +\delta^+_{A^\ast A'}
	\left\langle \hat{\sigma}_{A'}\left[\hat{O},
        \hat{\sigma}^\dagger_A\right]\right\rangle
	\right\} ,
\end{eqnarray}
where the notation $\sum_{A,A'}'$ indicates that \mbox{$A$ $\!\neq$ $A'$}.
In particular, when $\hat{O}$ is identified with the atomic operator
$\hat{\sigma}_A$, Eq.~(\ref{e3.1c}) yields
\begin{equation}
\label{e3.5}
       \langle\dot{\hat{\sigma}}_A \rangle =
       -{i\over \hbar}
       \left\langle\left[\hat{\sigma}_A,
       \hat{\tilde{H}}_{\rm S}\right]\right\rangle
       -\,i\sum_{\scriptstyle A' \atop \scriptstyle A'\neq A}
       \delta_{A^\ast A'}
       \left\langle\hat{\sigma}_{Az}\hat{\sigma}_{A'}\right\rangle .
\end{equation}
In Eqs.~(\ref{e3.1c}) and (\ref{e3.5}), the system Hamiltonian
$\,\hat{\!\tilde{H}}_{\rm S}$ is defined according to Eq.~(\ref{e3.1a}),
with $\omega_A$ being replaced by $\tilde{\omega}_A$, where
\begin{equation}
\label{e10}
    \tilde{\omega}_A = \omega_A-\delta_{A^\ast A},
\end{equation}
\begin{equation}
\label{e3.8}
       \delta_{A^\ast A} = \delta^-_{A^\ast A} -  \delta^+_{A^\ast A} ,
\end{equation}
\begin{equation}
\label{e3.8a}
       \delta^{-(+)}_{AA} = {{\cal P}\over \pi\hbar\varepsilon_0}
       \int_0^\infty {\rm d}\omega \,{\omega^2\over c^2}
       { {\bf d}_A \,{\rm Im}\,\bbox{G}({\bf r}_A,{\bf r}_A,\omega)\,{\bf d}_A
       \over \omega-(+)\omega_A}\,,
\end{equation}
and the resonant interatomic coupling parameters are defined
according to 
\begin{equation}
\label{e8}
       \left. \delta_{A^\ast A'}\right|_{A\neq A'}
       = \delta^-_{A^\ast A'} +  \delta^+_{A^\ast A'},
\end{equation}
\begin{eqnarray}
\label{e3.7}
\lefteqn{
       \left. \delta^{-(+)}_{AA'}\right|_{A\neq A'}
}
\nonumber\\&&\hspace{2ex}
       = {{\cal P}\over \pi\hbar\varepsilon_0}
       \int_0^\infty {\rm d}\omega \,{\omega^2\over c^2}
       { {\bf d}_A \,{\rm Im}\,\bbox{G} ({\bf r}_A,{\bf r}_{A'},\omega)
       \,{\bf d}_{A'}
       \over \omega-(+)\tilde{\omega}_{A'}}\,,
\end{eqnarray}
[${\cal P}$ -- principal value].
The notation $A^\ast$ ($A'^\ast$) means that ${\bf d}_A$ (${\bf d}_{A'}$)
in Eqs.~(\ref{e3.8a}) and (\ref{e3.7}) has to be replaced with
its complex conjugate ${\bf d}_A^\ast$ (${\bf d}_{A'}^\ast$).
Applying the Kramers-Kronig relation to the Green tensor, from
Eq.~(\ref{e3.7}) we derive that
\begin{equation}
\label{e10.0}
        \delta_{AA'}^-
        = {\tilde{\omega}_{A'}^2\over \hbar\varepsilon_0 c^2}\,
       {\bf d}_A\,
       {\rm Re}\,\bbox{G}({\bf r}_A,{\bf r}_{A'},
       \tilde{\omega}_{A'})\,{\bf d}_{A'} -
       \delta_{AA'}^+,
\end{equation}
and Eq.~(\ref{e8}) thus becomes
\begin{equation}
\label{e10.2}
       \left.\delta_{A^\ast A'}\right|_{A\neq A'} =
       {\tilde{\omega}_{A'}^2\over \hbar\varepsilon_0 c^2}\,
       {\bf d}_A^\ast {\rm Re}\,\bbox{G}({\bf r}_A,{\bf r}_{A'},
       \tilde{\omega}_{A'}){\bf d}_{A'} .
\end{equation}
Accordingly, from Eqs.~(\ref{e3.8a}) and (\ref{e3.8})
it follows that
\begin{equation}
\label{e10.1}
       \delta_{A^\ast A} =
       {\omega_A^2\over \hbar\varepsilon_0 c^2}\,
       {\bf d}_A^\ast {\rm Re}\,\bbox{G}({\bf r}_A,{\bf r}_A,
       \omega_{A}){\bf d}_A
       -  2\delta^+_{A^\ast A}.
\end{equation}
In dealing with problems of atoms embedded in media, the atoms should
be assumed to be localized in some small free-space regions,
so that the Green tensor at the positions of the atoms can always be
written as a sum of the vacuum Green tensor $\bbox{G}^V$
and the reflection Green tensor $\bbox{G}^R$.
Due to the singularity of ${\rm Re}\,\bbox{G}^V$
at equal space points, Eq. (\ref{e10.1}) actually applies to the
reflection part only. The vacuum
part, calculated in many textbooks, can be thought of
as being already included in $\omega_A$.

It should be pointed out that the single-atom
frequency shift $\delta_{A^\ast A}$ [Eq.~(\ref{e3.8}) or Eq.~(\ref{e10.1})]
(which is a real quantity  because of the reciprocity property of
the Green tensor) differs from the frequency shift $\delta_{A^\ast A}^-$
obtained in the rotating-wave approximation (see, e.g.,
Ref. \cite{Ho00}) in the term $\delta_{A^\ast A}^+$, which results
from the counter-rotating contributions to the Hamiltonian.
Note that when the atoms are in free space,
then Eqs. (\ref{e3.8}) and (\ref{e8}) reduce to the result derived in Refs.
\cite{Lehmberg70,Ackerhalt73,Agarwal74,Beige99,Skornia01}.
Since the first term on the right-hand side in Eq.~(\ref{e10.1})
can also be obtained classically \cite{Wylie85}, the second term
$\delta_{A^\ast A}^+$ is sometimes termed the quantum correction.
It is interesting to note that this quantum correction just
corrects the rotating-wave result. For that part of the
frequency shift which is caused by the presence of
the macroscopic bodies (mathematically, by the
reflection part
of the Green tensor), the quantum correction may safely be neglected.

{F}rom Eq.~(\ref{e10.2}) it is seen
that when the transition frequencies $\tilde{\omega}_{A}$ and
$\tilde{\omega}_{A'}$ of two atoms $A$ and $A'$ are different
from each other, then the strengths of the resonant dipole-dipole
coupling $|\delta_{A^\ast A'}|$ and $|\delta_{A'^\ast A}|$ are not
symmetric with respect to $A$ and $A'$. Only if the differences
\mbox{$|\tilde{\omega}_A$ $\!-$ $\!\tilde{\omega}_{A'}|$} are small
compared with the frequency scale of variation of the Green tensor, this
asymmetry can be disregarded and $\tilde{\omega}_A$ and
$\tilde{\omega}_{A'}$ may be replaced by an appropriately chosen
mid-frequency, say \mbox{$(\tilde{\omega}_A$ $\!+$
$\!\tilde{\omega}_{A'})/2$}. Whereas for
atoms in free space such an approximation is unproblematic
even for relatively large frequency differences, the
situation can drastically change if the presence of macroscopic
bodies gives rise to a highly peaked Green tensor that rapidly
varies with frequency.

Recalling the relationship
\begin{eqnarray}
\label{e10.2a}
      \bigl\langle \hat{O}(t) \bigr\rangle
      &=&
      {\rm Tr} \bigl[\hat{\rho}(0) \hat{O}(t)\bigr]
\nonumber\\[1ex] 
      &=& {\rm Tr} \bigl[\hat{\rho}(t) \hat{O}(0)\bigr]
      = {\rm Tr} \bigl[\hat{\varrho}(t) \hat{O}(0)\bigr],
\end{eqnarray}
where $\hat{O}$ is an arbitrary system operator, $\hat{\rho}$ is
the density operator of the overall system, and
$\hat{\varrho}$ is the (reduced) density operator of the
system, and making use of the cyclic properties of the trace,
from Eq.~(\ref{e3.1c}) we derive the following
equation of motion for the system density operator in
the Schr\"odinger picture (Appendix \ref{appA}):
\begin{eqnarray}
\label{e10.2d}
\lefteqn{
       \dot{\hat{\varrho}} =
       -{i\over \hbar} \left[\,\hat{\!\tilde{H}}_{\rm S},
       \hat{\varrho}\right]
}
\nonumber\\[1ex]&&\hspace{4ex}
       +\, \Biggl\{ i\sum_{A,A'}\sumprime
       \left[
       \delta^-_{A^\ast A'}
          \left(\hat{\sigma}^\dagger_A\hat{\sigma}_{A'}\hat{\varrho}
	  - \hat{\sigma}_{A'}\hat{\varrho}\hat{\sigma}^\dagger_A\right)
       \right.
\nonumber\\[1ex]&&\hspace{10ex}
        \left.
        +\,\delta^+_{A A^{'\ast}}
	   \left(\hat{\sigma}_A\hat{\sigma}^\dagger_{A'}\hat{\varrho}
	   -\hat{\sigma}^\dagger_{A'}\hat{\varrho}\hat{\sigma}_A\right)
        \right]
	+ {\rm H.c.}
	\Biggr\}.
\end{eqnarray}
In this equation both the resonant interatomic interaction
and the resonant atom-field interaction are taken into account,
without any restriction to the strength of the latter one.
It is worth noting that this type of equation cannot be
derived from an effective Hamiltonian in general.

In particular in the case when the resonant atom-field
interaction is sufficiently weak, it can also be treated
in a Markov approximation. Let $\hat{\varrho}$ be
the reduced density operator of the atomic system and let us
assume that the on-resonant part of the electromagnetic field
is initially also prepared in the vacuum state (i.e., there is
no external driving field). By tracing out both the on- and the
off- resonant medium-assisted field variables, we
arrive at the following master equation for the reduced
density operator of the atomic system (see Appendix \ref{appB}):
\begin{eqnarray}
\label{e17}
\lefteqn{
      \dot{\hat{\varrho}} =  - {\textstyle\frac{1}{2}}
      i \sum_A \tilde{\omega}_A\left[\hat{\sigma}_{Az},\hat{\varrho}\right]
}
\nonumber\\[1ex]&&\hspace{2ex}
      - \,{\textstyle{1\over 2}}
      \Biggl[ \sum_{A,A'}
      \Gamma_{A^\ast A'}
      \left(\hat{\sigma}^\dagger_A\hat{\sigma}_{A'}\hat{\varrho}
      - \hat{\sigma}_{A'}\hat{\varrho}\hat{\sigma}^\dagger_A\right)
	+ {\rm H.c.}
      \Biggr]
\nonumber\\[1ex]&&\hspace{2ex}
       + \,\Biggl\{ i\sum_{A,A'}\sumprime
       \left[
       \delta^-_{A^\ast A'}
       \left(\hat{\sigma}^\dagger_A\hat{\sigma}_{A'}\hat{\varrho}
       - \hat{\sigma}_{A'}\hat{\varrho}\hat{\sigma}^\dagger_A\right)
       \right.
\nonumber\\[1ex]&&\hspace{6ex}
        \left.
        +\,\delta^+_{A A^{'\ast}}
	\left(\hat{\sigma}_A\hat{\sigma}^\dagger_{A'}\hat{\varrho}
	-\hat{\sigma}^\dagger_{A'}\hat{\varrho}\hat{\sigma}_A\right)
        \right]
	+ {\rm H.c.}
	\Biggr\},
\end{eqnarray}
where
\begin{equation}
\label{e16}
\Gamma_{AA'} = {2\tilde{\omega}_{A'}^2\over \hbar\varepsilon_0c^2}\,
     {\bf d}_A \,{\rm Im}\,\bbox{G}({\bf r}_A,{\bf r}_{A'},
     \tilde{\omega}_{A'})\, {\bf d}_{A'}.
\end{equation}
If the differences between the atomic transition frequencies are small
in comparison with the frequency scale of variation of the
Green tensor, so that the relations
\begin{equation}
\label{e10.3}
\delta_{A^\ast A'}^\pm \simeq \delta_{A' A^\ast}^\pm
\end{equation}
and
\begin{equation}
\label{e18}
      \Gamma_{A^\ast A'} \simeq \Gamma_{A' A^\ast}
\end{equation}
are valid, then Eq.~(\ref{e17}) reduces to
\begin{eqnarray}
\label{e19}
\lefteqn{
      \dot{\hat{\varrho}} = - {\textstyle\frac{1}{2}}
      i \sum_A \tilde{\omega}_A
      \left[\hat{\sigma}_{Az},\hat{\varrho}\right]
      +i \sum_{A,A'}\sumprime
      \delta_{A^\ast A'}
      \left[\hat{\sigma}^\dagger_A \hat{\sigma}_{A'},
      \hat{\varrho}\right]
}
\nonumber\\[1ex]&&\hspace{2ex}
      - {\textstyle\frac{1}{2}}
      \sum_{A,A'}
      \Gamma_{A^\ast A'}
      \left( \hat{\sigma}^\dagger_A \hat{\sigma}_{A'} \hat{\varrho}
        -2 \hat{\sigma}_{A'} \hat{\varrho} \hat{\sigma}^\dagger_A
      +\hat{\varrho}\hat{\sigma}^\dagger_A \hat{\sigma}_{A'} \right)
      ,
\end{eqnarray}
which is of the same form as the master equation
obtained on the basis of Kubo's formula for the field correlation
functions \cite{Agarwal98}. 


\section{Effective Hamiltonian}
\label{effham}

Let us return to Eqs.~(\ref{e3.1c}) and (\ref{e10.2d}) and
restrict our attention to the case when the difference
between the atomic transition frequencies is small
in comparison with the frequency scale of variation of the
Green tensor, so that Eq.~(\ref{e10.3}) holds.
Then Eqs. (\ref{e3.1c}) and (\ref{e10.2d}) reduce to
\begin{equation}
\label{e10.3a}
       \bigl\langle \dot{\hat{O}} \bigr\rangle
       = -{i\over \hbar}
       \Bigl\langle \Bigl[\hat{O},
       \Bigl(
       \,\hat{\!\tilde{H}}_{\rm S}
       - \sum_{A,A'}\sumprime
       \hbar\delta_{A^\ast A'}
       \hat{\sigma}^\dagger_A\hat{\sigma}_{A'}
       \Bigr)\Bigr]\Bigr\rangle
\end{equation}
and
\begin{equation}
\label{e10.3b}
       \dot{\hat{\varrho}} =
       -{i\over \hbar}
       \Bigl[\Bigl(
       \,\hat{\!\tilde{H}}_{\rm S}
       - \sum_{A,A'}\sumprime
       \hbar\delta_{A^\ast A'}
       \hat{\sigma}^\dagger_A\hat{\sigma}_{A'}
       \Bigr)
       ,\hat{\varrho}\Bigr],
\end{equation}
respectively. Recalling the definition of $\,\hat{\!\tilde{H}}_S$,
we see that the motion of the system is governed
by the effective Hamiltonian
\begin{eqnarray}
\label{e11}
\lefteqn{
   \hat{H}_{\rm eff} = \int \!{\rm d}^3{\bf r}
   \Ints {\rm d}\omega \,\hbar\omega
   \,\hat{\bf f}^\dagger({\bf r},\omega){}\hat{\bf f}({\bf r},\omega)
}
\nonumber\\[1ex]&&\hspace{7ex}
    + \sum_{A} {\textstyle{1\over 2}}\hbar\tilde{\omega}_A \hat{\sigma}_{Az}
    - \sum_{A,A'}\sumprime 
    \hbar \delta_{A^\ast A'}
       \hat{\sigma}^\dagger_A\hat{\sigma}_{A'}
\nonumber\\[1ex]&&\hspace{7ex}
    - \sum_{A} \Ints {\rm d}\omega \left[
    \hat{\bf d}_A  \underline{\hat{\bf E}}({\bf r}_A,\omega)
    + {\rm H.c.}\right],
\end{eqnarray}
with $\underline{\hat{\bf E}}({\bf r}_A,\omega)$ being given
by Eq.~(\ref{e3}).

{F}rom the above it is clear that a Hamiltonian of the type (\ref{e1})
makes only sense if the atomic transition frequencies are sufficiently
near to each other. In that case, the Hamiltonian (\ref{e11})
is the desired extension of the Hamiltonian (\ref{e1}). The
Hamiltonian (\ref{e11}) is remarkable in several respects. Firstly,
it applies to atoms surrounded by arbitrarily configured,
dispersive and absorptive media. Secondly, it goes beyond the rotating-wave
approximation. Thirdly, it contains a resonant dipole-dipole coupling
energy that is explicitly expressed in terms of the
medium-assisted Green tensor, according to Eq.~(\ref{e10.2}).

The first term in Eq.~(\ref{e11}) describes, as before, the free
medium-assisted electromagnetic field energy. The second term is
the energy of the free atoms, which takes account of the
medium-induced single-atom transition frequency shift.
As already mentioned, the third term represents the energy of
the resonant dipole-dipole interaction of the atoms.
Note that both the single-atom
transition frequency shift and the interatom resonant dipole-dipole
interaction result from the off-resonant atom-field coupling.
The energy of the resonant interaction of the atoms with the
electromagnetic field is given by the fourth term,
which is of course not only responsible, e.g., for the single-atom
spontaneous decay in the regime of weak atom-field coupling and the
Rabi-type oscillations in the strong-coupling regime, but it
also determines, together with the resonant dipole-dipole coupling,
the mutual interaction of the atoms.
In particular, in the limit of a
$\delta$-like field excitation the atoms are resonantly
coupled to, the fourth term in Eq.~(\ref{e11}) reduces
to the interaction energy in the Tavis-Cummings model
\cite{Tavis68}. Note that within the Tavis-Cummings model
the resonant dipole-dipole coupling between the atoms
cannot be described, because this model does not
take account of the off-resonant atom-field interaction.
In practice the excitation spectrum of any real
electromagnetic field is always continuous, and the inclusion in
the Hamiltonian of the off-resonant atom-field interaction
is thus quite crucial, because it may lead to observable effects.

It is worth noting that, as can be seen from Eq.~(\ref{e10.2}),
the strength of the resonant dipole-dipole interaction is determined
by the real part of the Green tensor at different space points.
Thus, it is affected by a surrounding medium in a quite
different way as the spectral excitation density of the medium-assisted
electromagnetic field, which is determined by the imaginary part
of the Green tensor at equal space points. In particular, it may
happen that a high (low) excitation density and thus an enhanced (reduced)
single-atom decay rate is accompanied by a reduced (enhanced)
strength of the resonant dipole-dipole coupling \cite{Ho01b}.

Clearly, strong resonant dipole-dipole interaction can only be expected
if the atoms are sufficiently close to each other.
To give a simple example of the effect of material absorption, let us
assume that the atoms are embedded in bulk material of complex
permittivity $\varepsilon(\omega)$. Using the bulk-material
Green tensor (see, e.g., \cite{Knoll01}), from Eq.~(\ref{e10.2})
we find in the short-distance limit
\begin{equation}
\label{e12}
      \delta_{A^\ast A'} \!=\! {1 \over 4\pi\hbar\varepsilon_0 R^3}
      {\rm Re}\!\left[{1\over \varepsilon(\tilde{\omega}_{A'})}\right]\!\!
      \left( 3 \frac{{\bf d}_A^\ast {} {\bf R}}{R}\,
             \frac{{\bf d}_{A'} {} {\bf R}}{R}
             - {\bf d}_A^\ast {} {\bf d}_{A'}
      \!\right)
\end{equation}
(${\bf R}$ $\!=$ $\!{\bf r}_A$ $\!-$ $\!{\bf r}_{A'}$).
In free space, Eq.~(\ref{e12}) reduces to the well-known result
that the resonant dipole-dipole coupling simply corresponds to the
(near-field) Coulomb-type interaction. It is seen that the
characteristic $R^{-3}$ distance dependence observed in free space
is not changed by the medium.
In the long-distance limit, we find that
\begin{eqnarray}
\label{e12.1}
\lefteqn{
      \delta_{A^\ast A'} = {c^2
      \over 4\pi\hbar\varepsilon_0 R \tilde{\omega}_{A'}^2}
      \left( {\bf d}_A^\ast {} {\bf d}_{A'}
      - \frac{{\bf d}_A^\ast {} {\bf R}}{R}\, 
             \frac{{\bf d}_{A'} {} {\bf R}}{R} 
      \right) 
}
\nonumber\\[1ex]&&\hspace{3ex}\times\,
      \cos\!\left[n_{\rm R}(\tilde{\omega}_{A'})
      \frac{\tilde{\omega}_{A'}R}{c}\right]
      \exp\!\left[- n_{\rm I}(\tilde{\omega}_{A'})
      \frac{\tilde{\omega}_{A'}R}{c}\right]
\end{eqnarray}
[\mbox{$n(\omega)$ $\!=$
$\!n_{\rm R}(\omega)$ $\!+$ $\!in_{\rm I}(\omega)$
$\!=$ $\!\sqrt{\varepsilon(\omega)}]$}, i.e.,
the (harmonically modulated)
$R^{-1}$ dependence observed in free space
is changed to an exponential decrease according to
$\exp[-n_{\rm I}(\tilde{\omega}_{A'})\tilde{\omega}_{A'}R/c]$
due to material absorption. Thus, material absorption
can drastically reduce the resonant dipole-dipole coupling
strength with increasing mutual distance of the atoms.


\section{Temporal evolution of a two-atom system}
\label{sysdy}

Let us use the effective Hamiltonian (\ref{e11}) to study
a system of two-level atoms (resonantly) coupled to the
medium-assisted electromagnetic field and assume
that initially the atoms share a single excitation while the
field is in the vacuum state.
In the Schr\"{o}dinger picture we may write, on omitting
off-resonant terms, the state vector of the system in the form of
\begin{eqnarray}
\label{e4.1}
\lefteqn{
    |\psi(t)\rangle = \sum_A C_A(t)
    e^{-i(\tilde{\omega}_A-\bar{\omega})t}
    |U_A\rangle |\{0\}\rangle
}
\nonumber\\[1ex]&&\hspace{0ex}
     +\! \int \!{\rm d}^3{\bf r} \Ints \!\!{\rm d}\omega\,
     C_{Li}({\bf r},\omega,t)
     e^{-i (\omega-\bar{\omega})t}
     |L\rangle\hat{f}_i^\dagger({\bf r},\omega)|\{0\}\rangle
\end{eqnarray}
($\bar{\omega}$ $\!=$ $\!$ $\!\frac{1}{2}\sum_A\tilde{\omega}_A$).
Here, $|U_A\rangle$ is the atomic state with the $A$th atom
in the upper state and all the other atoms in the lower state,
and $|L\rangle$ is the atomic state with all atoms in the lower state.
Accordingly, $|\{0\}\rangle$ is the vacuum state of the
rest of the system, and
$\hat{f}^\dagger_i({\bf r},\omega) |\{0\}\rangle$
is the state, where a single quantum is excited.

{F}rom the Hamiltonian (\ref{e11}), the equations of motion for
the slowly varying probability amplitudes $C_A$ and ${\bf C}_L$
read as
\begin{eqnarray}
\label{e4.1a}
\lefteqn{
          \dot{C}_A(t) = \sum_{\scriptstyle A' \atop \scriptstyle A'\neq A}
	  i\delta_{A^\ast A'}
          e^{i(\tilde{\omega}_A-\tilde{\omega}_{A'})t} C_{A'}(t)
}
\nonumber \\[1ex]&&\hspace{6ex}
          -\frac{1}{\sqrt{\pi\epsilon_0\hbar}}
          \Ints \!\!{\rm d}\omega \,\frac{\omega^2}{c^2}
          \int {\rm d}^3{\bf r}\,\Big[
          \sqrt{\varepsilon_{\rm I}({\bf r},\omega)}
\nonumber \\[1ex]&&\hspace{10ex} \times\,
          {\bf d}_A^\ast \bbox{G}({\bf r}_A,{\bf r},\omega) \,
          {\bf C}_L({\bf r},\omega,t) e^{-i(\omega-\tilde{\omega}_A)t}\Big] ,
\end{eqnarray}
\begin{eqnarray}
\label{e4.1b}
\lefteqn{
          \dot{\bf C}_L({\bf r},\omega,t) =
          \frac{1}{\sqrt{\pi\epsilon_0\hbar}}\,
          \frac{\omega^2}{c^2}\,\sqrt{\varepsilon_{\rm I}({\bf r},\omega)}
}
\nonumber \\[1ex]&&\hspace{10ex} \times\,
          \sum_{A'}{\bf d}_{A'} \bbox{G}^\ast({\bf r}_{A'},{\bf r},\omega)\,
          C_{A'}(t) e^{i(\omega-\tilde{\omega}_{A'})t} .
\end{eqnarray}
By formally integrating Eq.~(\ref{e4.1b}) under the initial
condition that \mbox{${\bf C}_L(t=0)$ $\!=$ $\!0$}, and substituting
the formal solution into Eq.~(\ref{e4.1a}), we obtain the following
system of integrodifferential equations for the $C_A$:
\begin{eqnarray}
\label{e9}
\lefteqn{
        \dot{C}_A(t) =
	  \sum_{\scriptstyle A' \atop \scriptstyle A'\neq A}
	i\delta_{A^\ast A'}
        e^{i(\tilde{\omega}_A-\tilde{\omega}_{A'})t}
        C_{A'}(t)
}
\nonumber \\[1ex]&&\hspace{5ex}
        + \sum_{A'}
        \int_0^t {\rm d}t'
        \Ints \!{\rm d}\omega
         K_{A^\ast A'}(t,t';\omega)
        \, C_{A'}(t'),
\end{eqnarray}
where
\begin{eqnarray}
\label{e6}
\lefteqn{
        K_{AA'}(t,t';\omega)
        = -\frac{1} {\hbar\pi\varepsilon_0}
        \biggl[ {\omega^2\over c^2}
        e^{-i(\omega-\tilde{\omega}_A)t}
        e^{i(\omega-\tilde{\omega}_{A'})t'}
}
\nonumber \\[1ex]&&\hspace{20ex}\times
       {\bf d}_A {\rm Im}\,\bbox{G}({\bf r}_A,{\bf r}_{A'},\omega)
       {\bf d}_{A'} \biggr].
\end{eqnarray}
It should be pointed out that equations of the type
(\ref{e4.1}) -- (\ref{e6})
can also be derived on the basis of the original Hamiltonian
(\ref{e2}) in the rotating wave approximation
($\ints{\rm d}\omega\ldots$ $\!\to$
$\int_0^\infty{\rm d}\omega\ldots$), without the restrictive
condition that the Green tensor does not change on a frequency
scale defined by the differences of the atomic transition
frequencies \cite{Ho01a}. Clearly, in such an approach, the
contributions of the counter-rotating terms to the
frequency shifts and interatomic coupling strengths are disregarded.
In order to get insight into the atomic motion
on the basis of closed solutions of Eq.~(\ref{e9}),
let us consider two atoms and restrict our attention to
the limiting cases of weak and strong atom-field coupling.


\subsection{Weak atom-field coupling}
\label{weakcoupling}

In the weak coupling regime, the integral expression in
Eq.~(\ref{e9}) can be treated in a (coarse-grained) Markov
approximation, i.e., $C_{A'}(t')$ is replaced by $C_{A'}(t)$,
and the time integral $\int_0^t {\rm d}t'\,e^{-i(\omega
-\tilde{\omega}_{A'})(t-t')}$ is replaced by the
$\zeta$-function \mbox{$\zeta(\tilde{\omega}_{A'}$$\!-$ $\!\omega)$}.
Since the frequency integral only runs over the resonance
region, the $\zeta$-function effectively acts as the
$\delta$-function, and we arrive at the following
equations for the (slowly-varying) upper-state probability
amplitudes of two atoms $A$ and $B$:
\begin{eqnarray}
\label{e13}
       &&\dot{C}_A(t) = -{\textstyle\frac{1}{2}}
       \Gamma_{A^\ast A} C_A(t)
       +{\cal K}_{A^\ast B} e^{i(\tilde{\omega}_A-\tilde{\omega}_B)t}C_B(t),
\\[1ex] \label{e14}
       &&\dot{C}_B(t) = -{\textstyle\frac{1}{2}}
       \Gamma_{B^\ast B} C_B(t)
       +{\cal K}_{B^\ast A} e^{-i(\tilde{\omega}_A-\tilde{\omega}_B)t}C_A(t),
\end{eqnarray}
where
\begin{eqnarray}
\label{e15}
       {\cal K}_{AB}
       &=& - {\textstyle\frac{1}{2}} \Gamma_{AB} + i\delta_{AB}
\nonumber\\[1ex]
       &=& \,i {\tilde{\omega}_{B}^2\over \hbar\varepsilon_0c^2}\,
       {\bf d}_A \bbox{G}({\bf r}_A,{\bf r}_{B},
       \tilde{\omega}_{B}) {\bf d}_{B}\,.
\end{eqnarray}

We now make the simplifying assumption that the transition frequencies
of the two atoms are nearly equal to each other,
\mbox{$\tilde{\omega}_A$ $\!\simeq$ $\!\tilde{\omega}_B$}, but allow for
\mbox{$\Gamma_{A^\ast A}$ $\!\neq$ $\!\Gamma_{B^\ast B}$}.
The latter may happen, e.g., when the atoms
$A$ and $B$ have different dipole matrix elements and/or
different dipole orientations. It is then not difficult to
solve Eqs.~(\ref{e13}) and (\ref{e14}) analytically.
In particular, if the atom $B$ is initially in the lower state,
\mbox{$C_B(t$ $\!=$ $\!0)$ $\!=$ $\!0$}, we obtain
\begin{eqnarray}
\label{e16.1}
\lefteqn{
        C_A = \frac{1}{2D}\left\{
        \left[ - {\textstyle\frac{1}{2}}\left(
        \Gamma_{A^\ast A}-\Gamma_{B^\ast B}\right) + D \right]
        e^{D_+t/2}
        \right.
}
\nonumber\\[1ex]&&\hspace{12ex}
        \left.
        + \left[{\textstyle\frac{1}{2}}\left(
        \Gamma_{A^\ast A}-\Gamma_{B^\ast B}\right) + D \right] e^{D_-t/2}
        \right\},
\\[1ex]&&\hspace{0ex}
\label{e16.2}
        C_B = \frac{{\cal K}_{B^\ast A}}{D}
        \left(  e^{D_+t/2} - e^{D_-t/2}  \right),
\end{eqnarray}
where
\begin{equation}
\label{e16.3}
        D = \left[{\textstyle\frac{1}{4}}
        (\Gamma_{A^\ast A}-\Gamma_{B^\ast B})^2
        +4{\cal K}_{A^\ast B}{\cal K}_{B^\ast A}\right]^{1/2} ,
\end{equation}
\begin{equation}
\label{e16.4}
         D_\pm = -{\textstyle\frac{1}{2}}
         (\Gamma_{A^\ast A}+\Gamma_{B^\ast B}) \pm D .
\end{equation}

When the two atoms are identical and have equivalent positions
and dipole orientations with respect to the material
surroundings
such that the relations
\begin{eqnarray}
\label{e16.4a}
&\displaystyle
        \Gamma_{A^\ast A}=\Gamma_{B^\ast B},
\\
\label{e16.4b}
&\displaystyle
        \Gamma_{A^\ast B}=\Gamma_{B^\ast A},\quad
	\delta_{A^\ast B}=\delta_{B^\ast A}
\end{eqnarray}
are valid, we have \mbox{${\cal K}_{A^\ast B}$ $\!=$ $\!{\cal K}_{B^\ast A}$},
and $\Gamma_{A^\ast B}$, $\Gamma_{B^\ast A}$,
$\delta_{A^\ast B}$, and $\delta_{B^\ast A}$ are real quantities
due to the reciprocity of the Green tensor.
Then
Eqs.~(\ref{e16.1}) and (\ref{e16.2}) yield
the following expressions for the upper-state occupation
probabilities \mbox{$P_{A(B)}(t)$ $\!=|C_{A(B)}(t)|^2$}:
\begin{equation}
\label{e16.5}
         P_{A(B)}(t)= {\textstyle {1\over 2}}
         \left[ \cosh\left(\Gamma_{A^\ast B}t\right)
       +(-) \cos\left(2\delta_{A^\ast B}t\right) \right]
       e^{-\Gamma_{B^\ast B} t}.
\end{equation}

It is worth noting that Eqs.~(\ref{e16.1}) -- (\ref{e16.4})
[and thus Eq.~(\ref{e16.5})] are valid for arbitrary dispersing
and absorbing material surroundings of the atoms. In particular,
substituting in Eqs.~(\ref{e16}) and (\ref{e15}) for the Green tensor
the vacuum Green tensor, Eq.~(\ref{e16.5}) reduces to that one obtained
in Ref.~\cite{Lehmberg70}. Accordingly, using the Green tensor
for absorbing bulk material, the result in Ref.~\cite{Juzeliunas94}
is recognized.

{F}rom Eq.~(\ref{e16.5}) a damped oscillatory excitation
exchange between the two atoms is seen.
For sufficiently small times, \mbox{$\Gamma_{B^\ast B}t$
$\!\ll$ $\!1$}, and strong resonant dipole-dipole coupling,
\mbox{$|\delta_{A^\ast B}|\!$ $\!\gg \Gamma_{B^\ast B}$},
the oscillatory behavior
dominates. This can typically be observed when the atoms are
sufficiently near to each other, but can also be realized
for more moderate distances with the interatom coupling being
mediated by high-$Q$ medium-assisted field resonances
\cite{Ho01a}. In the opposite limit of weak resonant dipole-dipole
coupling, $P_A(t)$ decreases monotonously while $P_B(t)$ features
one peak which separates the regime of energy transfer from atom
$A$ to atom $B$ at early times and the subsequent decay of the
excited state of atom $B$.


\subsection{Strong atom-field coupling}
\label{strongcoupling}

For the sake of transparency, we again consider identical atoms
that have equivalent positions and dipole orientations
with respect to the material surroundings
so that Eqs. (\ref{e16.4a}) and (\ref{e16.4b}) hold.
Introducing the probability amplitudes
\begin{equation}
\label{e20}
      C_\pm (t) = 2^{-\frac{1}{2}}
      \left[ C_A(t) \pm C_B(t) \right]
      e^{\mp i\delta_{A^\ast B}t}
\end{equation}
of the superposition states
\begin{equation}
\label{e20a}
|\pm\rangle = 2^{-1/2} \left( |U_A\rangle\pm|U_B\rangle \right),
\end{equation}
from Eqs.~(\ref{e9}) we find that the equations for $C_+(t)$
and $C_-(t)$ decouple,
\begin{equation}
\label{e20.1}
        \dot{C}_{\pm}(t) =
        \int_0^t {\rm d}t' \Ints \!{\rm d}\omega \,
        K_\pm(t,t';\omega)\,e^{\mp i\delta_{A^\ast B}(t-t')}\, C_\pm(t'),
\end{equation}
where
\begin{equation}
\label{e20.2}
       K_\pm(t,t';\omega) = K_{A^\ast A}(t,t';\omega)
       \pm K_{A^\ast B}(t,t';\omega) .
\end{equation}

Let us restrict our attention to the case when
the absolute value of the two-atom term $K_{A^\ast B}(t,t';\omega)$
is of the same order of magnitude as the absolute value of the
single-atom term $K_{A^\ast A}(t,t';\omega)$, so that there is
a strong contrast in the magnitude of $K_+$ and $K_-$.
Typically true when the atoms are close to each other,
this may also take place at interatomic distances much larger
than the wavelength, e.g., as in the case of atoms
situated at diametrically opposite positions near a
microsphere \cite{Ho01a}.
As a consequence, the strong-coupling regime is applicable
to either the state $|+\rangle$ or the state
$|-\rangle$, but not to both at the same time.
Assuming that the field resonance strongly coupled to the
atoms has a Lorentzian shape,
with $\omega_m$
and $\Delta\omega_m$ being  the central frequency and
the half width at half maximum  respectively, we can
perform the frequency integral in Eq.~(\ref{e20.1}) in
a closed form, on extending it to $\pm\infty$,
\begin{eqnarray}
\label{e20.3}
\lefteqn{
        \Ints \!\!{\rm d}\omega\,K_\pm(t,t';\omega)
        \simeq - {\Gamma_\pm \over 2\pi} \,
        (\Delta\omega_m)^2
        e^{-i(\omega_m-\tilde{\omega}_A)(t-t')}
}
\nonumber\\[1ex]
        &&\hspace{15ex}\times\,
        \int_{-\infty}^{+\infty} \!\!{\rm d}\omega\,
        \frac{e^{-i(\omega-\omega_m)(t-t')}}
        {(\omega-\omega_m)^2 + (\Delta\omega_m)^2}
\nonumber\\[1ex]
        &&\hspace{3ex}
        = \,-\frac{\Gamma_\pm}{2} \Delta\omega_m
        e^{-i(\omega_m-\tilde{\omega}_A)(t-t')}
        e^{- \Delta\omega_m |t-t'|}\,,
\end{eqnarray}
where
\begin{equation}
\label{e20.4}
      \quad \Gamma_\pm = \Gamma_{A^\ast A}\pm\Gamma_{A^\ast B},
\end{equation}
with $\Gamma_{A^\ast A}$ and $\Gamma_{A^\ast B}$ being defined according
to Eq.~(\ref{e16}) with $\omega_m$ in place of
\mbox{$\tilde{\omega}_A$ $\!(\simeq$ $\!\tilde{\omega}_B)$}.
Substituting Eq. (\ref{e20.3}) into Eq. (\ref{e20.1}), and differentiating
both sides of the resulting equation with regard to time, we arrive at
\begin{eqnarray}
\label{e20.5}
\lefteqn{
         \ddot{C}_\pm(t)
         + \left[ i\left(\omega_m-\tilde{\omega}_A\pm
         \delta_{A^\ast B}\right)
	 + \Delta\omega_m \right] \dot{C}_\pm(t)
}
\nonumber\\[1ex]&&\hspace{25ex}
	 + \,\left(\Omega_\pm/2\right)^2 C_\pm(t) = 0,
\end{eqnarray}
where
\begin{equation}
\label{e22.0}
      \Omega_\pm = \sqrt{2\Gamma_\pm\Delta\omega_m}\ .
\end{equation}
In particular for exact resonance, i.e.,
\mbox{$\omega_m$ $\!=$ $\!\tilde{\omega}_A\mp\delta_{A^\ast B}$},
we derive
\begin{equation}
\label{e21}
       C_\pm(t) = 2^{-\frac{1}{2}}
       e^{-\Delta\omega_m t/2} \cos\!\left(\Omega_\pm t/2\right)
\end{equation}
($\Omega_\pm$ $\!\gg$ $\!\Delta\omega_m$). For the probability amplitudes
of the remaining states $|\mp\rangle$, which are weakly coupled to the field,
we obtain
\begin{equation}
\label{e22}
       C_\mp(t) = 2^{-\frac{1}{2}} e^{-\Gamma_\mp t/2}
\end{equation}
($\Omega_\mp$ $\!\ll$ $\!\Delta\omega_m$).
It then follows that
\begin{eqnarray}
\lefteqn{
       P_{A(B)}(t)=\textstyle{1\over 4} \Bigl[
       e^{-\Gamma_\mp t} + e^{-\Delta\omega_m t}
       \cos^2\!\left(\Omega_\pm t/2\right)
}
\nonumber\\[1ex]&&\hspace{2ex}
\label{e22.1}
     +(-) \,2e^{-(\Delta\omega_m  +\Gamma_\mp)t/2}
     \cos\!\left(\Omega_\pm t/2\right)
     \cos\!\left(2\delta_{A^\ast B}t\right) \Bigr].
\end{eqnarray}
Here the upper (lower) signs refer to the case where the state
$|+\rangle$ ($|-\rangle$) is strongly coupled to the medium-assisted
field.

Even if damping is ignored, $P_{A(B)}(t)$ is not
strictly periodic in general, because
$\Omega_\pm$ and $\delta_{A^\ast B}$ are not necessarily
commensurate with each other.
Roughly speaking, between three cases of approximately
periodic motion may be distinguished.
\begin{itemize}
\item[(i)] $4|\delta_{A^\ast B}|$ $\!\gg$ $\!\Omega_\pm$,
$t$ $\!\ll$ $\!2/\Omega_\pm$:
\begin{eqnarray}
\label{e23}
      P_A(t) &=& \cos^2(\delta_{A^\ast B}t),
\\
\label{e23.a}
      P_B(t) &=& \sin^2(\delta_{A^\ast B}t).
\end{eqnarray}
\item[(ii)] $4|\delta_{A^\ast B}|$ $\!\simeq$ $\!\Omega_\pm$,
$t$ $\!\ll$ $\!1/|2\delta_{A^\ast B}$ $\!-$ $\!\Omega_\pm/2|$:
\begin{eqnarray}
\label{e24}
      P_A(t) &=& \textstyle{1\over 4}
      \left[1+3\cos^2\!\left(\Omega_\pm t/2\right)\right],
\\
\label{e24.a}
      P_B(t) &=& \textstyle{1\over 4}
      \left[1-\cos^2\!\left(\Omega_\pm t/2\right)\right].
\end{eqnarray}
\item[(iii)] $4|\delta_{A^\ast B}|$ $\!\ll$ $\!\Omega_\pm$,
$t$ $\!\ll$ $\!1/|2\delta_{A^\ast B}|$:
\begin{eqnarray}
\label{e25}
      P_A(t) &=& \cos^4\!\left(\Omega_\pm t / 4\right),
\\
\label{e25.a}
      P_B(t) &=& \sin^4\!\left(\Omega_\pm t / 4\right).
\end{eqnarray}
\end{itemize}
{F}rom Eqs.~(\ref{e23}) -- (\ref{e25.a}) the following
time-averaged probabilities $\bar{P}_A$, $\bar{P}_B$,
and \mbox{$\bar{P}_L$ $\!=$ $\!1$ $\!-$ $\!\bar{P}_A$ $\!-$
$\!\bar{P}_B$} are obtained, with the respective time
integral being taken over one cycle.
(i) \mbox{$\bar{P}_A$ $\!=$ $\!\bar{P}_B$ $\!=$ ${1\over 2}$}
and \mbox{$\bar{P}_L$ $\!=$ $\!0$}. The excitation energy is
periodically exchanged between the two atoms through virtual
field excitations exclusively.
(ii) \mbox{$\bar{P}_A$ $\!=$ $\!{5\over 8}$}, \mbox{$\bar{P}_B$
$\!=$ ${1\over 8}$}, and \mbox{$\bar{P}_L$ $\!=$ $\!{2\over 8}$}.
The two exchange channels -- one channel through virtual and the other
one through real field excitations -- compete with each other and
destructively interfere, leading to a partial trapping of
the excitation energy in atom $A$. Note that this kind
of energy transfer suppression cannot be observed, if the
interference effect is disregarded as in Ref. \cite{Kurizki96}.
(iii) \mbox{$\bar{P}_A$ $\!=$ $\!\bar{P}_B$ $\!=$ ${3\over 8}$},
\mbox{$\bar{P}_L$ $\!=$ $\!{2\over 8}$}. This case is typically
observed in the long-distance limit, when the interatom energy
exchange is dominantly mediated by real field excitations.


\section{Rate regime}
\label{rateregime}

\begin{figure}[!t!]
\noindent
\begin{center}
\epsfig{figure=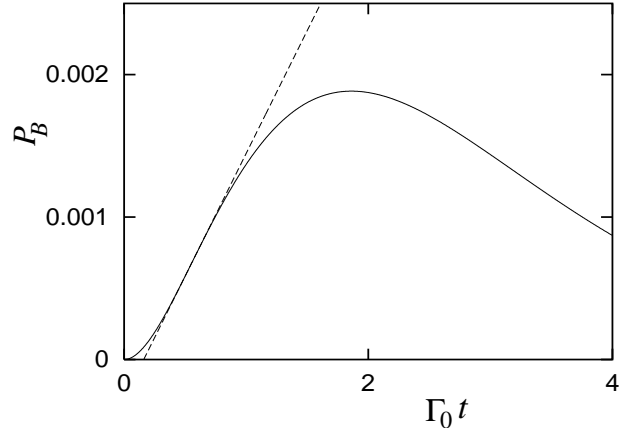,width=1\linewidth}
\end{center}
\caption{
The probability of finding atom $B$ in the upper state as a
function of time for two atoms situated at diametrically opposite
positions outside a dielectric microsphere.
The parameters have been chosen such that
$\Gamma_{A^\ast A}=\Gamma_{B^\ast B}=1.07\Gamma_0$,
$\Gamma_{A^\ast B}=0.04\Gamma_0$, and
$\delta_{A^\ast B}=0.06\Gamma_0$, 
$\Gamma_0$ being the single-atom decay rate in free space.
The solid curve is exact,
in accordance with Eq. (\ref{e16.5}), and the dashed
curve is $P_B(t) \simeq P_B(t_0)+w_1(t-t_0)$.
}
\label{rate}  
\end{figure}

Let us return to the case of weak atom-field coupling.
If the resonant dipole-dipole interaction is also weak,
then, as already mentioned in Subsection \ref{weakcoupling},
the energy transfer is one-way; that is, from atom $A$
to atom $B$. Clearly, the efficiency of such a transfer regime
is low, because only a small portion of energy is passed on
to atom $B$. In this case, a transfer rate $w_1$ can be 
introduced according to 
\begin{eqnarray}
\label{e16.5a}
      w_1={ {\rm d}P_B(t) \over {\rm d}t} \biggr|_{t_0},
\end{eqnarray} 
where $t_0$ is determined from the conditions that
\begin{eqnarray}
\label{e16.5b}
      { {\rm d}^2P_B(t) \over {\rm d}^2t} \biggr|_{t_0} = 0,
\qquad
      { {\rm d}P_B(t) \over {\rm d}t} \biggr|_{t_0} >0.
\end{eqnarray}
Note that this rate is not much different from that defined
as the ratio between the maximum value of $P_B(t)$ and the time
belonging to it \cite{Forster65}.
From Eq.~(\ref{e16.2}) [together with Eqs.~(\ref{e16.3}) and
(\ref{e16.4}) for negligibly small $K_{A^\ast B}$ and
$K_{B^\ast A}$ therein] it then follows that
\begin{eqnarray}
\label{e16.5c}
\lefteqn{
      w_1 \simeq {|K_{B^\ast A}|^2\over D^2} e^{D_-t_0}
}
\nonumber\\[1ex]&&\hspace{6ex}\times
      \left[ D_- + D_+ e^{2Dt_0}
      - (D_+ + D_-) e^{Dt_0} \right],
\end{eqnarray}
\begin{eqnarray}
\label{e16.5d}
\lefteqn{
      t_0 \simeq {1\over D}
      \ln\biggl( {1\over  4D_+^2} \Bigl\{(D_+ + D_-)^2
}       
\nonumber\\[1ex]&&\hspace{6ex}
      - \, 2D\left[(D_++D_-)^2 +4D_+D_-\right]^{1/2} \Bigr\}
     \biggr),
\end{eqnarray}
where
\begin{equation}
\label{e16.5e}
      D={\textstyle\frac{1}{2}}|\Gamma_{A^\ast A}-\Gamma_{B^\ast B}|,
\end{equation}
\begin{equation}
\label{e16.5f}
      D_{+(-)}=
      \left\{
      \begin{array}{l}
      -\Gamma_{B^\ast B(A^\ast A)} \quad{\rm if}\quad
        \Gamma_{A^\ast A}>\Gamma_{B^\ast B},\\[1ex]
      -\Gamma_{A^\ast A(B^\ast B)} \quad{\rm if}\quad
        \Gamma_{A^\ast A}<\Gamma_{B^\ast B}.
      \end{array}
      \right.
\end{equation}
Let us analyze Eq.~(\ref{e16.5c}) for three particular cases.
\begin{itemize}
\item[(i)]
If \mbox{$\Gamma_{A^\ast A}$ $\!\gg$ $\!\Gamma_{B^\ast B}$} is valid,
then from Eq.~(\ref{e16.5d}) it follows that
\mbox{$\Gamma_{A^\ast A}t_0$ $\!=$ $\!\ln 4$},
and Eq.~(\ref{e16.5c}) reduces to
\begin{equation}
\label{e16.7}
    w_1 = |{\cal K}_{B^\ast A}|^2/\Gamma_{A^\ast A}\,.
\end{equation}
\item[(ii)]
For \mbox{$\Gamma_{A^\ast A}$ $\!=$ $\!\Gamma_{B^\ast B}$}
the relation \mbox{$\Gamma_{B^\ast B}t_0$
$\!=$ $\!2$ $\!-$ $\!\sqrt{2}$} holds. Thus, Eq.~(\ref{e16.5c})
reads as
\begin{equation}
\label{e16.6}
    w_1 \simeq |{\cal K}_{B^\ast A}|^2 2\left(\sqrt{2}-1\right)
    e^{-\left(2-\sqrt{2}\right)}/\Gamma_{B^\ast B}\,.
\end{equation}
\item[(iii)]
In the case where \mbox{$\Gamma_{B^\ast B}$
$\!\gg$ $\!\Gamma_{A^\ast A}$} is valid, one finds
\mbox{$\Gamma_{B^\ast B}t_0$ $\!=$ $\!\ln 4$}, so that
Eq.~(\ref{e16.5c}) reduces to
\begin{equation}
\label{e16.7a}
        w_1=|{\cal K}_{B^\ast A}|^2/\Gamma_{B^\ast B}\,.
\end{equation}  
\end{itemize}
A typical example of the temporal evolution of $P_B(t)$ for weak
resonant dipole-dipole interaction is plotted in Fig.~\ref{rate}
for the case (ii). The rate regime with linear time dependence
is seen to be established after some short time interval where
a $t^2$-dependence is observed.

Another possible way for determining an energy transfer rate is to
infer it directly from the equations of motion. {F}rom Eq.~(\ref{e19})
[or Eqs.~(\ref{e13}) and (\ref{e14})] we find that
\begin{eqnarray}
\label{e16.7c}
       \dot{\varrho}_{BB} &=& -\Gamma_{B^\ast B} \varrho_{BB} 
               + ({\cal K}_{B^\ast A} \varrho_{AB} + {\rm c.c.}),
\\[1ex]
\label{e16.7d}
       \dot{\varrho}_{AB} &=& - {\textstyle\frac{1}{2}}
       \left(\Gamma_{A^\ast A}+\Gamma_{B^\ast B}\right) \varrho_{AB}
\nonumber\\[1ex]&&\hspace{4ex}        
               +\, {\cal K}_{A^\ast B} \varrho_{BB}
               + {\cal K}_{B^\ast A}^\ast \varrho_{AA}\,,
\end{eqnarray}
where the notation
\begin{equation}
\label{e16.7b}
       \varrho_{AA'} = 
       \langle U_A| \hat{\varrho} |U_{A'}\rangle = C_AC_{A'}^\ast 
\end{equation}
is used \mbox{($\tilde{\omega}_A$ $\!=$ $\!\tilde{\omega}_{A'}$)}.
In the rate regime \mbox{$\varrho_{BB}$ $\!(=$ $\!P_B)$}
varies (approximately) linearly with time. This may happen
for times where $\varrho_{AB}$ may be regarded as being
(qua\-si-)sta\-tionary.
By setting \mbox{$\dot{\varrho}_{AB}$ $\!=$ $\!0$} in
Eq.~(\ref{e16.7d}) and substituting the resulting expression for
$\varrho_{AB}$ into Eq.~(\ref{e16.7c}), we obtain
\begin{eqnarray}
\label{e16.7e}
\lefteqn{
       \dot{\varrho}_{BB} = -\left[\Gamma_{B^\ast B}
       -\left({2{\cal K}_{A^\ast B}{\cal K}_{B^\ast A} \over
          \Gamma_{AA^\ast }+\Gamma_{B^\ast B}} + {\rm c.c.}
          \right)\right] \varrho_{BB}
}
\nonumber\\[1ex]&&\hspace{20ex}
       +\, {4|{\cal K}_{B^\ast A}|^2 \over
          \Gamma_{A^\ast A}+\Gamma_{B^\ast B}} \, \varrho_{AA}\,.
\end{eqnarray}
In the rate regime, the small first term in Eq.~(\ref{e16.7e}) may
be neglected, and in the second term $\varrho_{AA}$ may be replaced
with an appropriately chosen \mbox{$P_A^{(0)}$ $\!=$
$\!\varrho_{AA}^{(0)}$}. The energy
transfer rate obtained in this way is then given by
\begin{equation}
\label{e16.7f}
      w_2 = {4|{\cal K}_{B^\ast A}|^2
      \over  \Gamma_{A^\ast A}+\Gamma_{B^\ast B}}\, P_A^{(0)}.
\end{equation}

The interest in energy transfer processes has stemmed from their key
role in a wide variety of both biological and nonbiological systems,
where the species participating in the energy transfer are typically
(not necessarily identical) molecules with dense manifolds of
vibronic states. Commonly, a rate regime is considered and the
transfer rate is calculated by means of Fermi's golden rule
in second-order perturbation theory with regard to the
molecule-field interaction,
\begin{eqnarray}
\label{e25.2a}
      w=\sum_{f,i}p_iw_{fi},
\end{eqnarray}
where $w_{fi}$ is proportional to the absolute square of the
second-order interaction matrix element and the energy conserving
$\delta$-function \mbox{$\delta(\omega_f$ $\!-$ $\!\omega_i)$}, and the
sum runs over the continuum of initial ($i$) and/or final ($f$) states.
In particular, short-distance energy transfer, where the intermolecular
coupling is essentially static, has been well known as F\"{o}rster
transfer \cite{Forster48}. Later on energy transfer over
arbitrary distances has been considered (see, e.g.,
Refs.~\cite{Juzeliunas94,Ho01b} and references therein). The
energy transfer rate in the presence of
dispersing and absorbing material surroundings
has been calculated in Ref.~\cite{Ho01b} on the basis
of a Hamiltonian of the type given in Eq.~(\ref{e2}).

For two two-level atoms the single-transition probability per unit time
$w_{if}$ corresponds to $\dot{P}_{B}$. Starting from the effective
Hamiltonian (\ref{e11}) and evaluating Eq.~(\ref{e9}) (for $C_{B}$) in
first-order perturbation theory, we derive,
on making the standard long-time assumption,
\begin{equation}
\label{e28}
       \dot{P}_B = 2\pi \left|{\cal K}_{B^\ast A}\right|^2
       \delta\!\left(\tilde{\omega}_A-\tilde{\omega}_B\right)
\end{equation}
(Appendix \ref{appC}), which is fully consistent with the result,
obtained in the standard second-order perturbation theory on the basis of the
original (fundamental) Hamiltonian (\ref{e2}) \cite{Ho01b}.
The only new feature is in the current treatment
the medium induced atomic frequency shifts are
taken into account. Note that the transition matrix element
is determined by the full Green tensor.

For establishing a rate regime, it is necessary that, according
to Eq.~(\ref{e25.2a}), a continuum of initial and/or final states
is involved in the transition. In the two-atom problem at hand,
the continua of states are obviously provided by the atomic level
broadening due to the spontaneous decay.
Let $\nu_A$ and $\nu_B$ be the frequencies of the continua
associated with atom $A$ and $B$ respectively and $\xi_A(\nu_A)$
and $\xi_B(\nu_B)$ the densities of the respective continua.
{F}rom Eqs.~(\ref{e25.2a}) and (\ref{e28}) the interatomic energy
transfer rate is then expected to be
\begin{eqnarray}
\label{e28.2}
\lefteqn{
     w= 2\pi\int {\rm d} \nu_A \int {\rm d} \nu_B \,
     \xi_A(\nu_A)\xi_B(\nu_B)
}
\nonumber\\[1ex]&&\hspace{10ex}\times\,
     p_A(\nu_A)
     |{\cal K}_{B^\ast A}|^2  \delta(\nu_A-\nu_B).
\end{eqnarray}
Assuming Lorentzian line shapes
\begin{equation}
\label{e28.3}
     \xi_{A'}(\nu_{A'}) = {1\over \pi} {\Gamma_{A'^\ast A'}/2\over
        (\nu_{A'}-\tilde{\omega}_{A'})^2 + (\Gamma_{A'^\ast A'}/2)^2}
\quad
    (A'=A,B),
\end{equation}
and making the assumption that $p_A(\nu_A)$ and
$|{\cal K}_{B^\ast A}|^2$ as functions of frequency slowly vary
over the lines, we may perform the integrals in Eq.~(\ref{e28.2})
to obtain
\begin{equation}
\label{e28.4}
      w = {4|{\cal K}_{B^\ast A}|^2 \over  \Gamma_{A^\ast A}
      +\Gamma_{B^\ast B}} \, p_A(\tilde{\omega}_A),
\end{equation}
which exactly agrees with the rate $w_2$ given by Eq.~(\ref{e16.7f}), if
the assumption is made that
\begin{equation}
\label{28.1}
     p_A(\tilde{\omega}_A)  \simeq P_A^{(0)}.
\end{equation}

Finally, let us identify $p_A(\tilde{\omega}_{A})$ in
Eq.~(\ref{e28.4}) for the perturbative rate $w$ with $P_A(t_0)$
[Eq.~(\ref{e22.1}) together with Eq.~(\ref{e16.5d})]
and compare the result with the nonperturbative rate $w_1$ calculated
from the exact temporal evolution of $P_B(t)$
[Eq.~(\ref{e16.5c}) together with Eq.~(\ref{e16.5d})].
Noting that \mbox{$P_A(t_0)$ $\!\simeq$
$\!\exp(-\Gamma_{A^\ast A}t_0)$}, we find for the three
cases considered in Eqs.~(\ref{e16.7}) -- (\ref{e16.7a})
the following results.
\begin{itemize}
\item[(i)]
\mbox{$\Gamma_{A^\ast A}$ $\!\gg$ $\!\Gamma_{B^\ast B}$}:
\begin{equation}
\label{e28.4a}
    P_A(t_0) \simeq {\textstyle\frac{1}{4}},
    \quad
    \frac{w_1}{w} \simeq 1.
\end{equation}
\item[(ii)]
\mbox{$\Gamma_{A^\ast A}$ $\!=$ $\!\Gamma_{B^\ast B}$}:
\begin{equation}
\label{e28.4b}
    P_A(t_0) \simeq e^{-\left(2-\sqrt{2}\right)},
    \quad
    \frac{w_1}{w} \simeq \sqrt{2}-1 \simeq 0.41.
\end{equation}
\item[(iii)]
\mbox{$\Gamma_{B^\ast B}$ $\!\gg$ $\!\Gamma_{A^\ast A}$}:
\begin{equation}
\label{e284c}
    P_A(t_0) \simeq 1,
    \quad
    \frac{w_1}{w} \simeq {\textstyle\frac{1}{4}}\,.
\end{equation}
\end{itemize}
Agreement between $w_1$ and $w$ is only observed in the first
case, where $\Gamma_{B^\ast B}$ is sufficiently small.
With increasing value of $\Gamma_{B^\ast B}$ an increasing
discrepancy between $w_1$ and $w$ is observed.

The reason can be seen in the fact that there are two processes
which simultaneously drive atom $B$: the energy transfer
from $A$ to $B$ and the spontaneous decay. Hence, $w_1$
is actually the rate of both processes combined,
not the ``naked'' rate of energy transfer. This explains why
$w_1$ is typically smaller than $w$ and why the discrepancy between
them becomes more substantial for enhanced
spontaneous decay of atom $B$. Note that the role of the spontaneous
decay of atom $B$ is twofold. On the one hand, it takes away
the population of the upper state of atom $B$, thus diminishes
$w_1$ relative to $w$. On the other hand, it leads to an atomic
level broadening which must carefully be taken into account for
a proper evaluation of the rate $w$ according to Eq.~(\ref{e28.2}).
To roughly compensate for the first effect,
the ratio $w_1/w$ may be multiplied by
$\exp(\Gamma_{B^\ast B}t_0)$. The values of the
so corrected ratio are then $\simeq 1$ for the cases
(i) and (iii), and $\simeq 0.74$ for the case (ii).


\section{Power spectrum}
\label{spectrum}

Let us finally address the question of the influence of the
resonant dipole-dipole interaction on the power spectrum of light
emitted by two identical atoms that initially share a single
excitation. As is well known, the (physical) spectrum can be
obtained by a Fourier transformation of the two-time correlation
function of the electric-field strength in normal order
(see, e.g., \cite{Vogel01})
\begin{eqnarray}
\label{e28.5}
\lefteqn{
          S({\bf r},\omega_{\rm S},T) =
          \int_0^T {\rm d}t_2\int_0^T {\rm d}t_1
	  \Bigl[e^{-i\omega_{\rm S}(t_2-t_1)}
}
\nonumber\\[1ex]&&\hspace{17ex}\times\,
          \left\langle \hat{\bf E}^{(-)}({\bf r},t_2)
          \hat{\bf E}^{(+)}({\bf r},t_1) \right\rangle\Bigr],
\end{eqnarray}
where $\omega_{\rm S}$ is the setting frequency of the (ideal)
spectral apparatus, $T$ is the operating-time interval of the detector,
and
\begin{equation}
\label{e28.6}
\hat{\bf E}^{(+)}({\bf r})
   = \int_{0}^{\infty} {\rm d}\omega\,
   \underline{\hat{\bf E}}({\bf r},\omega),
\end{equation}
\begin{equation}
\label{e28.6a}
\hat{\bf E}^{(-)}({\bf r})
   = \left[\hat{\bf E}^{(+)}({\bf r})\right]^\dagger.
\end{equation}


\subsection{Weak atom-field coupling}
\label{spectrum_wc}

For weak atom-field coupling we derive, on basing on
Eqs.~(\ref{e16.1}), (\ref{e16.2}), (\ref{e16.4a}), and (\ref{e16.4b}),
\begin{eqnarray}
\label{e18.1}
\lefteqn{
     S({\bf r},\omega_S,T \rightarrow \infty) = {1\over 4}
     \biggl|
     {{\bf F}_A+{\bf F}_B \over \Delta\omega_S
     + \delta_{A^\ast B} +i\Gamma_+/2}
}\qquad
\nonumber\\[1ex]&&\hspace{15ex}
     +\, {{\bf F}_A-{\bf F}_B\over \Delta\omega_S
     - \delta_{A^\ast B} +i\Gamma_-/2}
     \biggr|^2,
\end{eqnarray}
where
\begin{equation}
\label{e18.2}
           \Delta\omega_S=\omega_S-\tilde{\omega}_A
\end{equation}
and \mbox{($A'$ $\!=$ $\!A,B$)}
\begin{equation}
\label{e18.4}
          {\bf F}_{A'} =
          {\tilde{\omega}_{A'}^2 \over \pi\epsilon_0 c^2}
          \Ints {\rm d}\omega\,{\rm Im}\,\bbox{G}({\bf r},{\bf r}_{\rm A'},
          \omega)\,{\bf  d}_{A'}\zeta(\tilde{\omega}_{A'}-\omega)
\end{equation}
(Appendix \ref{appD}). Note that replacing \mbox{$\zeta(\tilde{\omega}_{A'}$
$\!-$ $\!\omega)$} with \mbox{$\pi\delta(\tilde{\omega}_{A'}$ $\!-$
$\!\omega)$} would be too rough here.
Equation (\ref{e18.1}) reveals that the
emitted light is spectrally split into two asymmetric lines
at \mbox{$\tilde{\omega}_A$ $\!\mp$ $\!\delta_{A^\ast B}$}, with $\Gamma_\pm$ and
\mbox{$|{\bf F}_A$ $\!\pm$ $\!{\bf F}_B|^2$} being the widths and weights
respectively. The line separation is seen to be twice the dipole-dipole
coupling parameter. A system of two two-level atoms, one of them
initially excited, is obviously equivalent to a three-level system
with two upper dressed states $|\pm\rangle$.
The doublet structure of the emitted-light spectrum can be easily understood
as a result of the transitions of the dressed states to the ground state.
For the case of the atoms being in vacuum, similar results were
found not only for the spontaneous emission \cite{Lehmberg70} but also
for the resonance fluorescence (see, e.g., Ref. \cite{Rudolph95} and
references therein).

As can be seen from Eq.~(\ref{e18.1}), an experimental observation
of doublet structure of the emitted light requires
a delicate balancing act. The interatomic distance should not be too large
to provide a reasonable level splitting, but it should not be too
small to avoid \mbox{$|{\bf F}_A|$ $\!=$ $\!|{\bf F}_B|$}, i.e.,
quenching of one of the two lines. It is worth noting that
the presence of macroscopic bodies may facilitate the detection
of the doublet, because it offers the possibility of realizing
strong resonant dipole-dipole interaction even for interatomic
distances much larger than the wavelength.

Another interesting feature
is that, according to Eq.~(\ref{e20.4}), either $\Gamma_+$ or
$\Gamma_-$, can be much smaller than \mbox{$\Gamma_{A^\ast A}$
$\!(=$ $\!\Gamma_{B^\ast B}$)}. That is, the resonant dipole-dipole
interaction can give rise to an ultranarrow spectral line, albeit
each time at the expense of the other line of the pair.
If, e.g., a single atom is placed sufficiently near a microsphere,
its spontaneous decay may be suppressed, with the emission line
being accordingly narrowed. Compared to the emission line of
a single atom, one line of the doublet observed
for two atoms being present may be further narrowed by
several orders of magnitude \cite{Ho01a}.


\subsection{Strong atom-field coupling}
\label{spectrum_sc}

For strong atom-field coupling, Eq.~(\ref{e18.1}) changes to,
on basing on Eqs.~(\ref{e21}) and (\ref{e22}),
\begin{eqnarray}
\label{e25.1}
\lefteqn{
     S({\bf r},\omega_S,T \rightarrow \infty)
     = {1\over 4}\biggl| ({\bf W}_A\pm {\bf W}_B)
}
\nonumber\\[1ex]&&\hspace{2ex}\,\times
     \biggl(
     {1\over  \Delta\omega_S\pm\delta_{A^\ast B}
      +\Omega_\pm/2+i\Delta\omega_m/2}
\nonumber\\[1ex]&&\hspace{4ex}
     -\,{1\over  \Delta\omega_S\pm\delta_{A^\ast B}
      -\Omega_\pm/2+i\Delta\omega_m/2}
     \biggr)
\nonumber\\[1ex]&&\hspace{8ex}
     +\,{i({\bf F}_A\mp {\bf F}_B) \over  \Delta\omega_S\mp \delta_{A^\ast B}
      +i\Gamma_\mp/2} \biggr|^2,
\end{eqnarray}
where \mbox{($A'$ $\!=$ $\!A,B$)}
\begin{equation}
\label{e25.2}
      {\bf W}_{A'}=
          {\omega_m^2\Delta\omega_m  \over
          \epsilon_0 c^2 \Omega_\pm}\,
          {\rm Im}\,\bbox{G}({\bf r},{\bf r}_{\rm A'},\omega_m)
          \,{\bf d}_{A'}
\end{equation}
(Appendix \ref{appD}).
Here the upper (lower) signs again refer to the case where
the state $|+\rangle$ ($|-\rangle$) is strongly coupled to the
medium-assisted field. Eq.~(\ref{e25.1}) reveals that
due to the strong atom-field coupling, the doublet observed for
weak atom-field coupling may become a triplet,
with one of the lines of the doublet being split into
two lines. These lines separated by $\Omega_\pm$
have equal widths (which are solely determined by the width of
the medium-assisted field resonance) and equal
weights. Note that their width and weight are different from those
of the third line, which is closely related to a line of
the doublet observed for weak atom-field coupling.
{F}rom Eq.~(\ref{e25.1}) it is also seen that, depending on the
point of observation,
this line or the strong-coupling-assisted doublet can be suppressed
due to the interference effects.


\section{Summary and concluding remarks}
\label{concl}

We have studied the interaction of two-level atoms with the
electromagnetic field in the presence of dispersing and absorbing
material surroundings described by a spatially varying permittivity
that is a complex function of frequency. Starting from the exact
multipolar Hamiltonian (in electric-dipole approximation), we have
derived reduced density-matrix equations of motion that describe
both the on-resonant interaction of the atoms with the medium-assisted
electromagnetic field and the resonant dipole-dipole interaction
of the atoms with each other via off-resonant atom-field
interaction. The equations of motion apply to atoms in
the presence of  arbitrarily configured, {\em dispersing} and
{\em absorbing} media and are not restricted to the  
rotating-wave approximation. Whereas the resonant dipole-dipole
interaction is essentially controlled by the real part of the
medium-assisted Green tensor, the relevant quantity for the
resonant atom-field interaction is the imaginary part of the
Green tensor. Both together determine the mutual interaction
of the atoms.  

We have shown that when the differences between the atomic
transition frequencies are small compared to the
frequency scale of variation of the medium-assisted Green tensor,
then the equations of motion can be derived from an
effective Hamiltonian. All coupling parameters are
again expressed in terms of the medium-assisted
Green tensor. The use of the effective Hamiltonian
may substantially simplify the calculations in cases where
identical atoms are considered.
Applying the theory to the two-atom case by assuming that initially the
medium-assisted electromagnetic field is in the ground state and one of the
atoms is in the upper state, we have considered the resonant
energy transfer between the atoms and the spectrum of the
emitted radiation.

There are two energy-transfer channels in general: one channel through
resonant dipole-dipole interaction mediated by virtual-photon creation
and destruction and the other one through emission
and absorption of real photons.
In particular for strong atom-field interaction,
the (over a period averaged) energy transfer
can be inhibited due to destructive interference of the two
available transfer channels.
When the dipole-dipole interaction is weak, the energy transfer from
the donor atom to the acceptor atom becomes irreversible, and
a transfer rate can be inferred from the (exact) temporal
evolution of the excited-state population of the acceptor atom.
Such a rate regime is commonly described by a transition probability
per unit time, such as the F\"{o}rster transfer rate, which is
calculated by means of Fermi's golden rule in second order
perturbation theory with respect to the fundamental
Hamiltonian. We have shown that the two methods essentially
lead to the same expression for the transfer rate. In any
case it is proportional to the absolute square of the
medium-assisted (full) Green tensor.
In contrast to molecules, where the necessary continuum of
initial/final states is typically given by the vibronic states,
in the case of atoms this continuum is essentially built up during
the process of spontaneous decay.

The results show that both the resonant energy transfer
and the doublet spectrum of the emitted light observed for
weak atom-field coupling or the triplet spectrum observed for
strong atom-field coupling can be
controlled by the presence of macroscopic bodies.
Clearly, the present analysis has left a number of open questions,
on which future work will concentrate. In particular, the
problem of energy exchange between atoms whose transition frequencies
must be regarded as being different with regard to the
variation of the medium-assisted Green tensor, needs
special emphasis.

\acknowledgements

This work was supported by the Deutsche Forschungsgemeinschaft.


\appendix

\newbox{\tmpbox}
\savebox{\tmpbox}{\bf Eqs.~(\ref{e3.1c}) and (\ref{e10.2d})}
\section{Derivation of \usebox{\tmpbox}}
\label{appA}

In order to perform the second term in Eq.~(\ref{e3.1b}),
\begin{eqnarray}
\label{A0}
\lefteqn{
       \hat{F}''(t) = {i\over \hbar} \sum_A \Intss {\rm d}\omega\,
       \left\{\left[\hat{O}(t),\hat{\bf d}_A(t)\right]
       \underline{\hat{\bf E}}({\bf  r}_A,\omega,t)
       \right.
}
\nonumber\\[1ex]&&\hspace{15ex}
       \left.
        +\,\underline{\hat{\bf E}}^\dagger({\bf  r}_A,\omega,t)
       \left[\hat{O}(t),\hat{\bf d}_A(t)\right]\right\},
\end{eqnarray}
we formally integrate Eq.~(\ref{e3.1}) to obtain
\begin{eqnarray}
\label{A1}
\lefteqn{
       \hat{\bf f}({\bf r},\omega,t)
       = \hat{\bf f}_{\rm free}({\bf r},\omega,t)
       + {\omega^2\over c^2}
        \sqrt{\varepsilon_{\rm I}({\bf r},\omega) \over  \hbar\pi\varepsilon_0}
}
\nonumber\\[1ex]&&\hspace{8ex}\times\;
        \sum_{A} \int_0^t {\rm d} t'\,
        \hat{\bf d}_{A}(t')\,\bbox{G}({\bf r}_{A},{\bf r},\omega)
        e^{-i\omega(t-t')},
\end{eqnarray}
where $\hat{\bf f}_{\rm free}({\bf r},\omega,t)$ evolves freely.
Inserting Eq.~(\ref{A1}) into Eq.~(\ref{e3}), we derive
\begin{eqnarray}
\label{A1.1}
\lefteqn{
       \underline{\hat{\bf E}}({\bf  r},\omega,t) =
       \underline{\hat{\bf E}}_{\rm free}({\bf  r},\omega,t)
}
\nonumber\\[1ex]&&\hspace{0ex}
        +\,{i \over \pi\varepsilon_0}{\omega^2\over c^2}
        \sum_A \int_0^t {\rm d} t'\,
        e^{-i\omega(t-t')}
        {\rm Im}\, \bbox{G}({\bf r},{\bf r}_A,\omega)
        \,\hat{\bf d}_A(t'),
\end{eqnarray}
where $\underline{\hat{\bf E}}_{\rm free}({\bf  r},\omega,t)$ is
defined according to Eq.~(\ref{e3}) with
$\hat{\bf f}_{\rm free}({\bf r},\omega,t)$ in place of
$\hat{\bf f}({\bf r},\omega,t)$.

Let us first restrict our attention to the single-atom case.
Introducing slowly varying atomic operator
\begin{equation}
\label{A2}
         \hat{\tilde{\sigma}}_A(t)
         = \hat{\sigma}_A(t) e^{i\omega_A t},
\end{equation}
we may rewrite Eq.~(\ref{e2.1}) as
\begin{equation}
\label{A2a}
        \hat{\bf d}_A(t)
        = {\bf d}_A\hat{\tilde{\sigma}}_A(t) e^{-i\omega_A t}
        + {\bf d}_A^\ast \hat{\tilde{\sigma}}_A^\dagger(t) e^{i\omega_A t}.
\end{equation}
Inserting Eqs.~(\ref{A1.1}) and (\ref{A2a}) into Eq.~(\ref{A0}),
we obtain
\begin{eqnarray}
\label{A3}
\lefteqn{
        \hat{F}''_A(t) = \hat{F}''_{A\,{\rm free}}
        -{1\over \hbar \pi\varepsilon_0}
        \!\int_0^t\!\!{\rm d}t'\!\! \Intss \!{\rm d}\omega\,
        {\omega^2\over c^2}
       \Bigl\{\left[\hat{O}(t),\hat{\bf d}_A(t)\right]
}
\nonumber \\[1ex]&&\hspace{1ex}\times\,
       {\rm Im}\,\bbox{G}({\bf r}_A,{\bf r}_A,\omega)
       \Bigl[
            {\bf d}_A \hat{\tilde{\sigma}}_A (t')
	    e^{-i(\omega-\omega_A)(t-t')}
            e^{-i\omega_At}
\nonumber \\[1ex]&&\hspace{8ex}
            +\,{\bf d}_A^\ast \hat{\tilde{\sigma}}_A^\dagger (t')
            e^{-i(\omega+\omega_A)(t-t')}
	    e^{i\omega_At} \Bigr]
\nonumber \\[1ex]&&\hspace{1ex}
       -\,{\rm Im}\,\bbox{G}({\bf r}_A,{\bf r}_A,\omega)
       \Bigl[
            {\bf d}_A \hat{\tilde{\sigma}}_A (t')
	    e^{i(\omega+\omega_A)(t-t')}
            e^{-i\omega_At}
\nonumber \\[1ex]&&\hspace{3ex}
            +\,{\bf d}_A^\ast \hat{\tilde{\sigma}}_A^\dagger (t')
            e^{i(\omega-\omega_A)(t-t')}
	    e^{i\omega_At}
       \Bigr]
       \left[\hat{O}(t),\hat{\bf d}_A(t)\right]
       \Bigr\},
\end{eqnarray}
where
\begin{eqnarray}
\label{A4.1}
\lefteqn{
       \hat{F}''_{A\,{\rm free}}(t) = {i\over \hbar} \Intss {\rm d}\omega\,
       \left\{\left[\hat{O}(t),\hat{\bf d}_A(t)\right]
       \underline{\hat{\bf E}}_{\rm free}({\bf  r}_A,\omega,t)
       \right.
}
\nonumber\\[1ex]&&\hspace{15ex}
       \left.
        +\,\underline{\hat{\bf E}}_{\rm free}^\dagger({\bf  r}_A,\omega,t)
       \left[\hat{O}(t),\hat{\bf d}_A(t)\right]\right\}.
\end{eqnarray}
Confining ourselves to resolving times large
compared with \mbox{$1/(\omega$ $\!-$ $\!\omega_A)$},
we may regard the exponential functions
in Eq.~(\ref{A3}) as being rapidly varying with time. Recall
that range of integration only covers off-resonant frequencies.
In the time integrals, we may then replace the slowly varying operator
$\hat{\tilde{\sigma}}_A(t')$ by $\hat{\tilde{\sigma}}_A(t)$
(Markov approximation). In the spirit of the coarse-grained time averaging
mentioned, we may further make the replacement
\begin{equation}
\label{A5}
       \int_0^t{\rm d}t'\, e^{-i(\omega-\omega_A)(t-t')}
        \rightarrow
       \zeta(\omega_A-\omega)
\end{equation}
[$\zeta(x)$ $\!=$ $\!\pi\delta(x)$ $\!+$ $\!i{\cal P}/x$]
and drop the terms where $\hat{\tilde{\sigma}}_A$
or $\hat{\tilde{\sigma}}_A^\dagger$ appears twice.
In this way, Eq.~(\ref{A3}) approximately reads as
\begin{eqnarray}
\label{A6}
\lefteqn{
        \hat{F}''_A(t) \simeq \hat{F}''_{A\,{\rm free}}(t)
}
\nonumber\\[1ex]&&\hspace{2ex}
        -\,{1\over \hbar \pi\varepsilon_0}
        \Intss \!\!{\rm d}\omega\, {\omega^2\over c^2}
       \,{\bf d}^\ast_A \,
       {\rm Im}\,\bbox{G}({\bf r}_A,{\bf r}_A,\omega)\,{\bf d}_A
\nonumber\\[1ex]&&\hspace{8ex}\times\,
       \Bigl\{
       \zeta[-(\omega+\omega_A)]
       \left[\hat{O}(t),\hat{\sigma}_A(t)\right]
       \hat{\sigma}^\dagger_{A}(t)
\nonumber\\[1ex]&&\hspace{12ex}
       +\,\zeta(\omega_A-\omega)
       \left[\hat{O}(t),\hat{\sigma}^\dagger_A(t)\right]
       \hat{\sigma}_{A}(t)
\nonumber\\[1ex]&&\hspace{8ex}
       -\,\zeta(\omega+\omega_A)
       \,\hat{\sigma}_{A}(t)\left[\hat{O}(t),
       \hat{\sigma}^\dagger_A(t)\right]
\nonumber\\[1ex]&&\hspace{12ex}
       -\,\zeta(\omega-\omega_A)
       \,\hat{\sigma}^\dagger_{A}(t)\left[\hat{O}(t),
       \hat{\sigma}_A(t)\right]
       \Bigr\}.
\end{eqnarray}
Substituting Eq.~(\ref{A6}) into Eq.~(\ref{e3.1b}) and taking
into account that, because of \mbox{$\omega$ $\!\pm$
$\omega_A$ $\!\neq$ $\!0$}, the $\delta$-function parts of
the $\zeta$-functions do not contribute to the
frequency integrals, we make the approximation
\begin{equation}
\label{A6a}
\zeta(\omega \pm \omega_A)
\rightarrow i{\cal P}/(\omega \pm \omega_A)
\end{equation}
and arrive at
\begin{eqnarray}
\label{A7}
       \,\dot{\!\hat{O}}
       = -{i\over \hbar}
       \left[\hat{O},\,\hat{\!\tilde{H}}_{A\,{\rm S}}\right]
       + \hat{F}''_{A\,{\rm free}}\,.
\end{eqnarray}
Here, $\,\hat{\!\tilde{H}}_{A\,{\rm S}}$ is the single-atom
system Hamiltonian defined according to Eq.~(\ref{e3.1a}),
with the ``naked'' transition frequency $\omega_A$ being replaced by
the shifted frequency \mbox{$\tilde{\omega}_A$ $\!=$
$\!\omega_A$ $\!-$ $\!\delta_{A^\ast A}$}, where the frequency shift
$\delta_{A^\ast A}$ is given by Eq.~(\ref{e3.8}) together with
Eq.~(\ref{e3.8a}).

Let us turn to the multi-atom case. Substituting the
formal solution (\ref{A1.1}) into Eq.~(\ref{A0}), we now obtain
a double-sum over $A$ and $A'$,
\begin{equation}
\label{A7a}
\hat{F}''(t) = \sum_A \hat{F}''_A(t)
+ \sum_{A,A'}\sumprime \hat{F}''_{AA'}(t).
\end{equation}
The terms $\hat{F}''_A(t)$
are treated as described above to give the shifted
transition frequency for each atom. In the terms
$\hat{F}''_{AA'}(t)$
\mbox{($A$ $\!\neq$ $\!A'$)} we then take into account the
level shifts and introduce the slowly varying
atomic operators according to
\begin{eqnarray}
\label{A8}
         \hat{\tilde{\sigma}}_A(t)
         = \hat{\sigma}_A(t) e^{i\tilde{\omega}_A t}.
\end{eqnarray}
By repeating for the terms $\hat{F}''_{AA'}(t)$
the same procedure as in the single-atom case, it is not difficult
to prove that [in place of Eq.~(\ref{A6})] the result is
\begin{eqnarray}
\label{A8a}
\lefteqn{
        \hat{F}''_{AA'}(t) \simeq
        -\,{1\over \hbar \pi\varepsilon_0}
        \Intss \!\!{\rm d}\omega\, {\omega^2\over c^2}
        \,\Bigl\{
       {\bf d}_A \,{\rm Im}\,\bbox{G}({\bf r}_A,
       {\bf r}_{A'},\omega)\,{\bf d}^\ast_{A'}
}
\nonumber\\[1ex]&&\hspace{15ex}\times\,
       \zeta[-(\omega\!+\!\tilde{\omega}_{A'})]
       \left[\hat{O}(t),\hat{\sigma}_A(t)\right]
       \!\hat{\sigma}^\dagger_{A'}(t)
\nonumber\\[1ex]&&\hspace{0ex}
       +\,{\bf d}^\ast_A \,{\rm Im}\,\bbox{G}({\bf r}_A,
       {\bf r}_{A'},\omega)\,{\bf d}_{A'}
       \zeta(\tilde{\omega}_{A'}\!-\!\omega)
       \!\left[\hat{O}(t),\hat{\sigma}^\dagger_A(t)\right]
       \!\hat{\sigma}_{A'}(t)
\nonumber\\[1ex]&&\hspace{0ex}
       -\,{\bf d}^\ast_A \,{\rm Im}\,\bbox{G}({\bf r}_A,
       {\bf r}_{A'},\omega)\,{\bf d}_{A'}
       \zeta(\omega\!+\!\tilde{\omega}_{A'})
       \hat{\sigma}_{A'}(t)\!\left[\hat{O}(t),
       \hat{\sigma}^\dagger_A(t)\right]
\nonumber\\[1ex]&&\hspace{0ex}
       -\,{\bf d}_A \,{\rm Im}\,\bbox{G}({\bf r}_A,
       {\bf r}_{A'},\omega)\,{\bf d}^\ast_{A'}
       \zeta(\omega\!-\!\tilde{\omega}_{A'})
       \hat{\sigma}^\dagger_{A'}(t)\!\left[\hat{O}(t),
       \hat{\sigma}_A(t)\right]
       \!\Bigr\}.
\nonumber\\&&
\end{eqnarray}
In Eq.~(\ref{A8a}) we now make the approximation
\begin{equation}
\label{A8b}
\zeta(\omega \pm \tilde{\omega}_{A'})
\rightarrow i{\cal P}/(\omega \pm \tilde{\omega}_{A'}),
\end{equation}
[cf. Eq.~(\ref{A6a})] and insert the resulting expression for
$\hat{F}''_{AA'}(t)$ into Eq.~(\ref{A7a}). We eventually combine
Eqs.~(\ref{e3.1b}) and (\ref{A7a}) and obtain, on again dropping
off-resonant terms,
\begin{eqnarray}
\label{A9}
\lefteqn{
       \,\dot{\!\hat{O}}
       = -{i\over \hbar}
       \left[\hat{O},\,\hat{\!\tilde{H}}_{\rm S}\right] + \hat{F}''_{\rm free}
}
\nonumber\\[1ex]&&\hspace{2ex}
       +\,i\sum_{A,A'}\sumprime
       \Bigl\{
       \delta^-_{A^\ast A'}
       \left[\hat{O},\hat{\sigma}^\dagger_A\right]\hat{\sigma}_{A'}
        +\delta^+_{A A^{'\ast}}
       \left[\hat{O},\hat{\sigma}_A\right]\hat{\sigma}^\dagger_{A'}
\nonumber\\[1ex]&&\hspace{6ex}
        +\,\delta^-_{A A^{'\ast}}
	\hat{\sigma}^\dagger_{A'}\left[\hat{O},\hat{\sigma}_A\right]
        +\delta^+_{A^\ast A'}
	\hat{\sigma}_{A'}\left[\hat{O},\hat{\sigma}^\dagger_A\right]
	\Bigr\},
\end{eqnarray}
where
\begin{equation}
\label{A9aa}
        \hat{F}''_{\rm free}= \sum_A \hat{F}''_{A\,{\rm free}}\,,
\end{equation}
and $\delta_{AA'}$ is given by Eq.~(\ref{e8}) together with
Eq.~(\ref{e3.7}). From Eq.~(\ref{A4.1}) it is seen that when
the off-resonant free field is in the vacuum state, then
\begin{equation}
\label{A9a}
\bigl\langle\hat{F}''_{\rm free}\bigr\rangle =0
\end{equation}
is valid, and the expectation value of the operator equation (\ref{A9})
just yields Eq.~(\ref{e3.1c}).

To prove Eq.~(\ref{e10.2d}), we use the cyclic properties
of the trace and rewrite Eq.~(\ref{e3.1c}) as
\begin{eqnarray}
\label{A10}
\lefteqn{
      \frac{{\rm d}}{{\rm d}t}\bigl\langle\hat{O}\bigr\rangle
      = {\rm Tr}
      \Biggl( \biggl\{
      -{i\over \hbar} \left[\,\hat{\!\tilde{H}}_{\rm S},\hat{\varrho}\right]
}
\nonumber\\[1ex]&&\hspace{2ex}
       +\,i\sum_{A,A'}\sumprime
       \Bigl[
       \delta^-_{A^\ast A'}
               \left(\hat{\sigma}^\dagger_A\hat{\sigma}_{A'}\hat{\varrho}
	       - \hat{\sigma}_{A'}\hat{\varrho}\hat{\sigma}^\dagger_A\right)
\nonumber\\[1ex]&&\hspace{10ex}
        +\,\delta^+_{A A^{'\ast}}
	       \left(\hat{\sigma}_A\hat{\sigma}^\dagger_{A'}\hat{\varrho}
	       -\hat{\sigma}^\dagger_{A'}\hat{\varrho}\hat{\sigma}_A\right)
\nonumber\\[1ex]&&\hspace{10ex}
        +\,\delta^-_{A A^{'\ast}}
	       \left(\hat{\sigma}_A\hat{\varrho}\hat{\sigma}^\dagger_{A'}
	       -\hat{\varrho}\hat{\sigma}^\dagger_{A'}\hat{\sigma}_A\right)
\nonumber\\[1ex]&&\hspace{10ex}
        +\,\delta^+_{A^\ast A'}
	       \left(\hat{\sigma}^\dagger_A\hat{\varrho}\hat{\sigma}_{A'}
	       -\hat{\varrho}\hat{\sigma}_{A'}\hat{\sigma}^\dagger_A\right)
        \Bigr]
	\biggr\} \hat{O}\Biggr).
\end{eqnarray}
The relationship (\ref{e10.2a}) obviously implies the relationship
\begin{equation}
\label{A10a}
{\rm Tr}\,\left[\frac{{\rm d}\hat{\varrho}(t)}{{\rm d}t}\,\hat{O}(0)\right]
= {\rm Tr}\,\left[\hat{\varrho}(0)\,\frac{{\rm d}\hat{O}(t)}{{\rm d}t}\right].
\end{equation}
Since $\hat{O}$ is an arbitrary system operator, from Eqs.~(\ref{A10})
and (\ref{A10a}) it then follows that Eq.~(\ref{e10.2d}) must be valid. 


\newbox{\tmpbox}
\savebox{\tmpbox}{\bf Eq.~(\ref{e17})}
\section{Derivation of \usebox{\tmpbox}}
\label{appB}
 
For weak atom-field coupling, the on-resonant term
\begin{eqnarray}
\label{B0}
\lefteqn{
       \hat{F}'(t) = {i\over \hbar} \sum_A \Ints {\rm d}\omega\,
       \left\{\left[\hat{O}(t),\hat{\bf d}_A(t)\right]
       \underline{\hat{\bf E}}({\bf  r}_A,\omega,t)
       \right.
}
\nonumber\\[1ex]&&\hspace{15ex}
       \left.
        +\,\underline{\hat{\bf E}}^\dagger({\bf  r}_A,\omega,t)
       \left[\hat{O}(t),\hat{\bf d}_A(t)\right]\right\},
\end{eqnarray}
in the Hamiltonian $\,\hat{\!\tilde{H}}_S$ in Eq.~(\ref{A9})
can also be treated within the
approximation scheme outlined in Appendix \ref{appA} for
the off-resonant term $\hat{F}''(t)$ [Eq.~(\ref{A0})].
Writing $\hat{F}'(t)$ as
\begin{equation}
\label{B0.1}
       \hat{F}'(t) = \hat{F}'_{\rm free}+\sum_{A,A'} F'_{AA'}(t)
\end{equation}
where $\hat{F}'_{\rm free}$ is defined according to Eqs. (\ref{A4.1})
and (\ref{A9aa}), but with $\ints{\rm d}\omega\ldots$ in
place of $\intss{\rm d}\omega\ldots$,
and repeating the steps leading to Eq.~(\ref{A8a}),
we obviously arrive at
an equation for $\hat{F}'_{AA'}(t)$ which again looks like
Eq.~(\ref{A8a}), but again with $\ints{\rm d}\omega\ldots$
in place of $\intss{\rm d}\omega\ldots$.
Note that the equation for $\hat{F}'_{AA'}(t)$ also applies
to the case \mbox{$A$ $\!=$ $\!A'$}. Since the frequency integrals
now run over the resonance region,
only the $\delta$-function parts
of the $\zeta$-functions do effectively contribute to the
frequency integrals, i.e., we may approximate the $\zeta$-functions
by $\delta$-functions,
\begin{equation}
\label{BB4.1}
\zeta(\omega \pm \tilde{\omega}_{A'})
\rightarrow \pi\delta(\omega \pm \tilde{\omega}_{A'}).
\end{equation}
Inserting Eq.~(\ref{B0.1}) [together with the equation for
$\hat{F}'_{AA'}(t)$ and Eq.~(\ref{BB4.1})] into the expression
for $\,\hat{\!\tilde{H}}_{\rm S}$ in the first term on the right-hand
side of Eq.~(\ref{A9}), we derive
\begin{eqnarray}
\label{BB5}
\lefteqn{
       \,\dot{\!\hat{O}} = -{\textstyle{1\over 2}}i\sum_A
       \tilde{\omega}_A\left[\hat{O}, \hat{\sigma}_{Az}\right]
       + \hat{F}_{\rm free}
}
\nonumber\\[1ex]&&\hspace{2ex}
       -{\textstyle{1\over 2}}\sum_{A,A'}
       \biggl\{
       \Gamma_{A^\ast A'}
       \left[\hat{O},\hat{\sigma}^\dagger_A\right]\hat{\sigma}_{A'}
	+ \Gamma_{A A^{'\ast}}
	\hat{\sigma}^\dagger_{A'}\left[\hat{O},\hat{\sigma}_A\right]
	\biggr\}
\nonumber\\[1ex]&&\hspace{2ex}
       +\,i\sum_{A,A'}\sumprime
       \biggl\{
       \delta^-_{A^\ast A'}
       \left[\hat{O},\hat{\sigma}^\dagger_A\right]\hat{\sigma}_{A'}
        +\delta^+_{A A^{'\ast}}
       \left[\hat{O},\hat{\sigma}_A\right]\hat{\sigma}^\dagger_{A'}
\nonumber\\[1ex]&&\hspace{6ex}
        +\,\delta^-_{A A^{'\ast}}
	\hat{\sigma}^\dagger_{A'}\left[\hat{O},\hat{\sigma}_A\right]
        +\delta^+_{A^\ast A'}
	\hat{\sigma}_{A'}\left[\hat{O},\hat{\sigma}^\dagger_A\right]
	\biggr\},
\end{eqnarray}
where $\Gamma_{AA'}$ is given by Eq.~(\ref{e16}), and
\begin{equation}
\label{BB5.1}
\hat{F}_{\rm free} = \hat{F}'_{\rm free} + \hat{F}''_{\rm free}.
\end{equation}
Assuming that the medium-assisted electromagnetic free field is
in the vacuum state, thus
\begin{equation}
\label{BB5.2}
\bigl\langle\hat{F}_{\rm free}\bigr\rangle = 0,
\end{equation}
and recalling the procedure that has led from Eq.~(\ref{A9}) 
via (\ref{A10}) to Eq.~(\ref{e10.2d}), we see
that Eq.~(\ref{BB5}) just leads to Eq. (\ref{e17}).


\newbox{\tmpbox}
\savebox{\tmpbox}{\bf Eq.~(\ref{e28})}
\section{Derivation of \usebox{\tmpbox}}
\label{appC}

To calculate \mbox{$P_B(t)$ $\!=$ $\!|C_B(t)|^2$} in lowest-order of
perturbation theory with regard to Eq.~(\ref{e9}),
we replace $C_A$ and $C_B$ on the right-hand side of the equation
for $\dot{C}_B$ by their respective initial values $1$ and $0$,
and integrate both sides of the equation with respect to time to obtain
\begin{eqnarray}
\label{e26}
\lefteqn{
      C_B(t) = 
      i\delta_{B^\ast A} \int_0^t {\rm d}t'
      \,e^{i(\tilde{\omega}_B-\tilde{\omega}_A)t'}
}
\nonumber\\[1ex]&&\hspace{5ex}
      + \int_0^t\!{\rm d}t' \int_0^{t'}\!{\rm d}t''
      \Ints \!\!{\rm d}\omega
      K_{B^\ast A}(t',t'';\omega).
\end{eqnarray}
Using Eq.~(\ref{e6}) together with the relationship
\begin{eqnarray}
\label{B1}
\lefteqn{
       e^{-i\omega(t'-t'')}\theta(t'-t'')
}
\nonumber\\[1ex]&&\hspace{5ex}
       =\, -{1\over 2\pi i}
       \int_{-\infty}^\infty {\rm d}\omega'
       \,{e^{-i\omega'(t'-t'')} \over \omega' - \omega +i0_+}\,,
\end{eqnarray}
we may rewrite Eq.~(\ref{e26}) as
\begin{eqnarray}
\label{B2}
\lefteqn{
      C_B(t) =
      i\delta_{B^\ast A} 2\pi
      \delta^{(t)}(\tilde{\omega}_A-\tilde{\omega}_B)
}
\nonumber\\[1ex]&&\hspace{2ex}
      +\,{2\over i \hbar\varepsilon_0}
       \!\int_{-\infty}^\infty \!{\rm d}\omega'
       \!\Ints \!{\rm d}\omega\, {\omega^2\over c^2}
       {{\bf d}_B^\ast \,{\rm Im}\,\bbox{G}({\bf r}_B,{\bf r}_A,\omega)
       \, {\bf d}_A
       \over \omega' - \omega +i0_+} \quad
\nonumber\\[1ex]&& \hspace{10ex}\times\,
        \delta^{(t)}(\omega'-\tilde{\omega}_B)
	\,\delta^{(t)}(\tilde{\omega}_A-\omega'),
\end{eqnarray}
where
\begin{equation}
\label{B2a}
       \delta^{(t)}(\omega) =
       {1\over 2\pi}
       \int_0^t {\rm d}t'\, e^{-i\omega t'} .
\end{equation}
The function \mbox{$\Delta(\omega')$ $\!=$
$\!\delta^{(t)}(\omega'$ $\!-$ $\!\tilde{\omega}_B)
\,\delta^{(t)}(\tilde{\omega}_A$ $\!-$ $\!\omega')$}
in Eq. (\ref{B2}) is essentially different from zero in an
interval around \mbox{$\tilde{\omega}_A$ $\!\simeq$
$\!\tilde{\omega}_B$}, the extension of which is of the order
of magnitude of $1/t$.
For sufficiently long times $t$, it is reasonable to assume that
the function
\begin{equation}
\label{B2b}
    R(\omega') = \Ints \!{\rm d}\omega\, {\omega^2\over c^2}
       {{\bf d}_B^\ast \,{\rm Im}\,\bbox{G}({\bf r}_B,{\bf r}_A,\omega)
       \, {\bf d}_A
       \over \omega' - \omega +i0_+}
\end{equation}
is slowly varying on the frequency scale of variation of
$\Delta(\omega')$.
The integral over $\omega'$ can then be performed separately
to yield
\begin{equation}
\label{B2c}
       \int_{-\infty}^\infty \!{\rm d}\omega'\,
        \delta^{(t)}(\omega'-\tilde{\omega}_B)\,
	\delta^{(t)}(\tilde{\omega}_A-\omega')
        = \delta^{(t)}(\tilde{\omega}_A-\tilde{\omega}_B),
\end{equation}
and Eq.~(\ref{B2}) takes the form of \mbox{($t$ $\!\to$ $\!\infty$)}
\begin{eqnarray}
\label{B3}
\lefteqn{
      C_B(t) =
      \biggl[ 2\pi i\delta_{B^\ast A}
      +{2\over  i\hbar\varepsilon_0}\,
}
\nonumber\\[1ex]&&\hspace{0ex}\times\!
       \Ints \!\!{\rm d}\omega {\omega^2\over c^2}
       {
       {\bf d}_B^\ast \,{\rm Im}\,\bbox{G}({\bf r}_B,{\bf r}_A,\omega)
       \,{\bf d}_A
       \over
       \tilde{\omega}_A - \omega +i0_+ }
       \biggr]
      \delta^{(t)}(\tilde{\omega}_A-\tilde{\omega}_B)
       .
\end{eqnarray}
To evaluate the frequency integral in Eq.~(\ref{B3}), we complete
it by adding the corresponding off-resonant part and subsequently
subtract it. The integral $\int_0^{\infty}{\rm d}\omega\ldots$
can then be approximately evaluated
by extending the lower limit to $-\infty$
and using contour integral technique in a similar way as
in Ref.~\cite{Ho01b}. The remaining integral
$\intss{\rm d}\omega\ldots$ can again be treated as a principal-value
integral to give $-2\pi i\delta_{B^\ast A}^-$
[cf. Eq.~(\ref{e3.7})], which
(apart from the quantum correction) cancels out
the first term in Eq.~(\ref{B3}).
Taking into account that for \mbox{$t$ $\!\to$ $\!\infty$},
$|\delta^{(t)}(\omega)|^2$ behaves like
\begin{equation}
\label{B5}
       \bigl|
         \delta^{(t)}(\omega)\bigr|^2 =
       {t\over 2\pi}\,\delta(\omega),
\end{equation}
we eventually arrive at
\begin{eqnarray}
\label{B6}
\lefteqn{
       P_B(t)=|C_B(t)|^2 =
      {2\pi\over \hbar^2} \left({\tilde{\omega}_A^2\over
      \varepsilon_0 c^2}\right)^2
}
\nonumber\\[1ex]&&\hspace{5ex}\times\,
      |{\bf d}_B^\ast \,\bbox{G} ({\bf r}_B,{\bf  r}_A,\tilde{\omega}_A)
      \,{\bf d}_A|^2
      \delta(\tilde{\omega}_A-\tilde{\omega}_B) t ,
\end{eqnarray}
which together with Eq.~(\ref{e15}) leads to Eq.~(\ref{e28}).


\newbox{\tmpbox}
\savebox{\tmpbox}{\bf Eqs.~(\ref{e18.1}) and (\ref{e25.1})}
\section{Derivation of \usebox{\tmpbox}}
\label{appD}

Let us briefly outline the derivation of Eqs.~(\ref{e18.1})
and (\ref{e25.1}) (for more details, see \cite{Ho00}).
In the Schr\"odinger picture, the two-time correlation function
in Eq.~(\ref{e28.5}) can be given by
\begin{eqnarray}
\label{D1}
\lefteqn{
          \left\langle \hat{\bf E}^{(-)}({\bf r},t_2)
          \hat{\bf E}^{(+)}({\bf r},t_1) \right\rangle
}
\nonumber\\[1ex]&&\hspace{0ex}
          = \left\langle \psi(t_2) \left|
	  \hat{\bf E}^{(-)}({\bf r})
	  e^{-i\hat{H}_{\rm eff}(t_2-t_1)/\hbar}
          \hat{\bf E}^{(+)}({\bf r}) \right|\psi(t_1)\right\rangle.
\end{eqnarray}
Calculating it by using Eqs.~(\ref{e4.1}), (\ref{e4.1b}),
(\ref{e28.6}), (\ref{e28.6a}), and ({\ref{e3}),
and inserting the resulting expression into Eq.~(\ref{e28.5}),
we derive  
\begin{eqnarray}
\label{D2}
\lefteqn{
          S({\bf r},\omega_{\rm S},T) =
           \Biggl|
          \sum_{A} \int_0^T \!\!{\rm d}t_1\biggl[
          e^{i(\omega_{\rm S}-\tilde{\omega}_{A}) t_1}
          \!\!\int_0^{t_1}\!\! {\rm d}t'\,C_{A}(t')
}
\nonumber\\[1ex]&&\hspace{1ex}\times
          \Ints
          \!\!{\rm d}\omega\,{\omega^2\over \pi\epsilon_0 c^2}\,
          {\rm Im}\,\bbox{G} ({\bf r},{\bf r}_{A},\omega) {\bf d}_{A}
          e^{-i(\omega-\tilde{\omega}_{A}) (t_1-t')}
          \biggr]\Biggr|^2 . 
\nonumber\\
\end{eqnarray}
For the two-atom system under consideration,
in the weak-coupling regime,
$C_{A(B)}(t')$ may be replaced with $C_{A(B)}(t_1)$, where $C_{A(B)}(t_1)$
is given according to Eqs.~(\ref{e16.1}) and (\ref{e16.2}),
\begin{eqnarray}
\label{D3}
\lefteqn{
      C_{A(B)} (t) = {\textstyle\frac{1}{2}}
}
\nonumber\\[1ex]&&\hspace{2ex}\times\,
      \left[ e^{(-\Gamma_+/2+i\delta_{A^\ast B})t}
      +(-) e^{(-\Gamma_-/2-i\delta_{A^\ast B})t} \right].
\end{eqnarray}
(\mbox{$\Gamma_{A^\ast A}$ $\!=$ $\!\Gamma_{B^\ast B}$}
and \mbox{${\cal K}_{A^\ast B}$ $\!=$ $\!{\cal K}_{B^\ast A}$}).
The $t'$-integral may then be regarded, in the long-time limit,
as being \mbox{$\zeta(\tilde{\omega}_{A'}$ $\!-$ $\!\omega)$}, so that the
factor ${\bf F}_{A'}$ [Eq.~(\ref{e18.4})] can be put in front of the
$t_1$-integral. Now the $t_1$-integral can be performed to
obtain Eq.~(\ref{e18.1}) \mbox{($T$ $\!\to$ $\!\infty$)}.

For strong atom-field coupling, we use Eq.~(\ref{e20}) and
express in Eq.~(\ref{D2}) $C_{A(B)}(t)$ in terms of $C_\pm(t)$,
\begin{eqnarray}
\label{D4}
\lefteqn{
          S({\bf r},\omega_{\rm S},T) =
          {1\over 2} \biggl|
          \int_0^T {\rm d}t_1
          e^{i(\omega_{\rm S}-\tilde{\omega}_A) t_1}
}
\nonumber\\[1ex]&&\hspace{1ex}\times\,
          \biggl\{ \int_0^{t_1} {\rm d}t'\,
          C_+(t')e^{i\delta_{A^\ast B}t'}
	  \Ints {\rm d}\omega\,{\omega^2\over \pi\epsilon_0 c^2}\,
\nonumber\\[1ex]&&\hspace{7ex}\times\,
          \left[{\rm Im}\,\bbox{G} ({\bf r},{\bf r}_A,\omega) \,{\bf d}_A
          + {\rm Im}\,\bbox{G} ({\bf r},{\bf r}_B,\omega) \,{\bf d}_B\right]
\nonumber\\[1ex]&&\hspace{15ex}\times\,
          e^{-i(\omega-\tilde{\omega}_A) (t_1-t')}
\nonumber\\&&\hspace{2ex}
          +\, \int_0^{t_1} {\rm d}t'\,
          C_-(t')e^{-i\delta_{A^\ast B}t'}
	  \Ints {\rm d}\omega\,{\omega^2\over \pi\epsilon_0 c^2}\,
\nonumber\\[1ex]&&\hspace{7ex}\times\,
          \left[{\rm Im}\,\bbox{G} ({\bf r},{\bf r}_A,\omega) \,{\bf d}_A
          - {\rm Im}\,\bbox{G} ({\bf r},{\bf r}_B,\omega) \,{\bf d}_B\right]
\nonumber\\[1ex]&&\hspace{15ex}\times\,
          e^{-i(\omega-\tilde{\omega}_A) (t_1-t')}
          \biggr\}\biggr|^2 .
\end{eqnarray}
Recall that either the state $|+\rangle$ or the state $|-\rangle$
is strongly coupled to the medium-assisted electromagnetic field,
but not both at the same time. The state which is weakly coupled to
the field can be treated in the same way as above. For the strongly
coupled state, we again assume a
Lorentzian shape for the field resonance 
($\omega_m$, central frequency; $\Delta\omega_m$, width;
cf. Section \ref{strongcoupling}) and evaluate the
$\omega$-integral. Taking $C_\pm(t')$ from Eq.~(\ref{e21}),
we then can evaluate the remaining time integrals to obtain
Eq.~(\ref{e25.1}) together with Eq.~(\ref{e25.2})
\mbox{($T$ $\!\to$ $\!\infty$)}.



\begin{references}
\bibitem[*]{byline} On leave from the Institute of Physics, National Center
for Natural Sciences and Technology, 1 Mac Dinh Chi Street,
District 1, Ho Chi Minh city, Vietnam.

\bibitem{Lukin00}
M. D. Lukin and P. R. Hemmer, Phys. Rev. Lett. {\bf 84}, 2818 (2000).

\bibitem{Barenco95}
A. Barenco, D. Deutsch, A. Ekert, and R. Jozsa, Phys. Rev. Lett. 
{\bf 74}, 4083 (1995).

\bibitem{John91}
S. John and J. Wang, Phys. Rev. B {\bf 43}, 12772 (1991);
G. Kweon and N. M. Lawandy, J. Mod. Opt. {\bf 41}, 311 (1994);
S. John and T. Quang,  Phys. Rev. A {\bf 52}, 4083 (1995);
S. Bay, P. Lambropoulos, and K. M{\o}lmer, {\it ibid.} {\bf 55}, 1485 (1997).

\bibitem{Kobayashi95} 
T. Kobayashi, Q. Zheng, and T. Sekiguchi, Phys. Rev. A {\bf 52}, 2835 (1995).

\bibitem{Takada99}
A. Takada and K. Ujihara, Opt. Commun. {\bf 160}, 146 (1999).

\bibitem{Juzeliunas94} 
G. Juzeli\=unas and D. L. Andrews, Phys. Rev. B {\bf 50}, 13371 (1994). 

\bibitem{Agarwal98}
G. S. Agarwal and S. D. Gupta, Phys. Rev. A {\bf 57}, 667 (1998).

\bibitem{Ho01a}
Ho Trung Dung, S. Scheel, L. Kn\"oll, and D.-G. Welsch, 
J. Opt. B: Quant. Semiclass. Opt. {\bf 4}, S169 (2002).

\bibitem{Goldstein97}
E. V. Goldstein and P. Meystre, Phys. Rev. A {\bf 56}, 5135 (1997).

\bibitem{Kurizki88}
G. Kurizki and A. Z. Genack, Phys. Rev. Lett. {\bf 61}, 2269 (1988).

\bibitem{Zheng96}
Q. Zheng, T. Kobayashi, and T. Sekiguchi, 
Phys. Rev. Lett. {\bf 77}, 406 (1996);
G. Kurizki, A. G. Kofman, and A. Z. Genack,
{\em ibid.} {\bf 77}, 407 (1996).

\bibitem{Kurizki96}
G. Kurizki, A. G. Kofman, and V. Yudson, Phys. Rev. A {\bf 53}, R35 (1996).

\bibitem{Knoll01}
L. Kn\"{o}ll, S. Scheel, and D.-G. Welsch,
in {\it Coherence and Statistics of Photons and Atoms},
edited by J. Pe\v{r}ina (John Wiley \& Son, New York, 2001), p. 1.

\bibitem{Ho01b}
Ho Trung Dung, L. Kn\"oll, and D.-G. Welsch,
Phys. Rev. A {\bf 65}, 043813 (2002).

\bibitem{Ho00}
S. Scheel, L. Kn\"{o}ll, and D.-G. Welsch,  
Phys. Rev. A {\bf 60}, 4094 (1999);
Ho Trung Dung, L. Kn\"oll, and D.-G. Welsch, 
{\it ibid.} {\bf 62}, 053804 (2000).

\bibitem{Lehmberg70}
R. H. Lehmberg, Phys. Rev. A {\bf 2}, 889 (1970).

\bibitem{Ackerhalt73}
J. R. Ackerhalt, P. L. Knight, and J. H. Eberly, Phys. Rev. Lett.
{\bf 30}, 456 (1973);
J. R. Ackerhalt and J. H. Eberly, Phys. Rev. D {\bf 10}, 3350 (1974).

\bibitem{Agarwal74}
G. S. Agarwal, {\it Quantum Optics}, Vol. 70 of 
{\it Springer Tracts in Modern Physics} (Springer, Berlin, 1974).

\bibitem{Beige99}
A. Beige and G. C. Hegerfeldt, Phys. Rev. A {\bf 59}, 2385 (1999).

\bibitem{Skornia01}
C. Skornia, J. von Zanthier, G. S. Agarwal, E. Werner, and H. Walther,
Phys. Rev. A {\bf 64}, 053803 (2001).

\bibitem{Wylie85}
J. M. Wylie and J. E. Sipe, Phys. Rev. A {\bf 32}, 2030 (1985);
J. P. Dowling, Found. Phys. {\bf 28}, 855 (1998).

\bibitem{Tavis68}
M. Tavis and F. W. Cummings, Phys. Rev. {\bf 170}, 379 (1968).

\bibitem{Forster65}
T. F\"orster, in {\it Modern Quantum Chemistry}, 
edited by O. Sinagoglu (Academic, New York, 1965).

\bibitem{Forster48}
T. F\"orster, Ann. Phys. (Leipzig) {\bf 1}, 55 (1948); 
D. L. Dexter, J. Chem. Phys. {\bf 21}, 836 (1953).

\bibitem{Vogel01}
W. Vogel, D.-G. Welsch, and S. Wallentowitz, 
{\it Quantum Optics, An Introduction}
(Wiley-VCH, Berlin, 2001).

\bibitem{Rudolph95}
T. G. Rudolph, Z. Ficek, and B. J. Dalton, Phys. Rev. A {\bf 52}, 636 (1995).

\end{references}
\end{document}